\documentclass[12pt]{article}

\usepackage{amsmath}
\usepackage{graphicx}
\usepackage{natbib}
\usepackage{url} % not crucial - just used below for the URL 
\usepackage{xcolor}
\usepackage{amssymb}
\usepackage{booktabs}
\usepackage{multirow}
\usepackage{rotating}
\usepackage{array, longtable, makecell}
\usepackage{caption}
\usepackage{calrsfs}
\usepackage[lofdepth,lotdepth]{subfig}
\usepackage[title]{appendix}
\usepackage{enumitem}
\usepackage{mathtools}

\usepackage{float}
\floatstyle{plaintop}
\restylefloat{table}
\usepackage{subfig}

\def\bX{\boldsymbol{X}}
\def\bY{\boldsymbol{Y}}
\def\bM{\boldsymbol{M}}
\def\bU{\boldsymbol{U}}
\def\b0{\boldsymbol{0}}
\def\bw{\boldsymbol{w}}
\def\bz{\boldsymbol{z}}
\def\bB{\boldsymbol{B}}

\def\bbeta{\boldsymbol{\beta}}
\def\bTheta{\boldsymbol{\Theta}}

\def\bphi{\boldsymbol{\phi}}
\def\bPhi{\boldsymbol{\Phi}}
\def\bPsi{\boldsymbol{\Psi}}

\newcommand{\Sa}{\mathcal{S}}
\newcommand{\real}{\mathbb{R}}

\DeclareMathOperator{\E}{\mathbb{E}}
\DeclareMathOperator{\tr}{\text{tr}}
\DeclareMathOperator{\vect}{vec}

\DeclareMathAlphabet{\pazocal}{OMS}{zplm}{m}{n}

\title{Matrix Normal Cluster-Weighted Models}
\author{Salvatore D. Tomarchio$^1$ \and Paul D. McNicholas$^2$ \and Antonio Punzo$^1$}
\date{%
    $^1$Department of Economics and Business, University of Catania\\%
    $^2$Department of Mathematics and Statistics, McMaster University\\[2ex]%
}

\begin{document}
\maketitle

\def\spacingset#1{\renewcommand{\baselinestretch}%
{#1}\small\normalsize} \spacingset{1}

%%%%%%%%%%%%%%%%%%%%%%%%%%%%%%%%%%%%%%%%%%%%%%%%%%%%%%%%%%%%%%%%%%%%%%%%%%%%%%

\bigskip
\begin{abstract}

Finite mixtures of regressions with fixed covariates are a commonly used model-based clustering methodology to deal with regression data.
However, they assume assignment independence, i.e.~the allocation of data points to the clusters is made independently of the distribution of the covariates.
In order to take into account the latter aspect, finite mixtures of regressions with random covariates, also known as cluster-weighted models (CWMs), have been proposed in the univariate and multivariate literature.
In this paper, the CWM is extended to matrix data, e.g.~those data where a set of variables are simultaneously observed at different time points or locations.
Specifically, the cluster-specific marginal distribution of the covariates, and the cluster-specific conditional distribution of the responses given the covariates, are assumed to be matrix normal.
Maximum likelihood parameter estimates are derived using an ECM algorithm.
Parameter recovery, classification assessment and the capability of the BIC to detect the underlying groups are analyzed on simulated data.
Finally, two real data applications concerning educational indicators and the Italian non-life insurance market are presented. 

\end{abstract}

\noindent%
{\it Keywords:} Mixture models, Matrix-variate, Classification, Clustering.
\vfill

\newpage
\spacingset{1.5} % DON'T change the spacing!
\section{Introduction}
\label{sec:intro}

Finite mixture models have seen increasing use over the last decades \citep[for a recent survey, see][]{mcnicholas16b}.
Because of their flexibility, they are a suitable statistical tool for modeling a wide range of phenomena characterized by unobserved heterogeneity, and constitute a powerful device for clustering and classification.
Specifically, in their ``direct application'', each mixture component represents a group (or cluster) within data, and the scope is to identify these groups and estimate the parameters of the conditional-group distributions \citep{mclachlan2000finite, mcnicholas2016mixture}.
If no exogenous variables explain the means and the variances of each component, they are also called unconditional mixture models.
However, when there is a linear relationship between some variables, important insight can be gained by accounting for functional dependencies between them.
For this reason, finite mixtures of regression models with fixed covariates (FMR) have been proposed in the literature \citep[see][for examples]{desarbo1988, fruhwirth2006finite}.
Finite mixtures of regression models with concomitant variables (FMRC; \citealp{dayton1988concomitant}) are an extension of FMR where the mixing weights depend on some concomitant variables (which are often the same covariates) and are usually modeled by a multinomial logistic model (see \citealp{ingrassia2016decision, ingrassia2019cluster} and \citealp{Mazz:Batt:Ingr:Punz:Mode:2019} for details).
Unfortunately, these methodologies do not explicitly use the distribution of the covariates for clustering, i.e.~the assignment of data points to clusters does not directly utilize any information from the distribution of the covariates.
Differently from these approaches, finite mixtures of regressions with random covariates (\citealp{gershenfeld1997, gershenfeld1999}), also known as cluster-weighted models (CWMs), allow for such functional dependency.
This occurs because, for each mixture component, CWMs decompose the joint distribution of responses and covariates into the product between the marginal distribution of the covariates and the conditional distribution of the responses given the covariates.

Several CWMs have been introduced in the univariate and multivariate literature.
Most of them consider a univariate response variable, along with a set of covariates, modeled by a univariate and a multivariate distribution, respectively \citep[see][for examples]{ingrassia2012local, ingrassia2014model, punzo2014flexible}.
Instead, fewer CWMs exist in the case of a multivariate response \citep{punzo2017robust, dang2017multivariate}.
However, over the years, there has been an increasing interest in applications involving matrix-variate (three-way) data. 
This data structure can occur in several and different application domains, such as longitudinal data on multiple response variables, spatial multivariate data, multivariate repeated measures or spatio-temporal data.
Nevertheless, there exists a limited number of contributions involving matrix-variate regression models.
First introduced by \citet{viroli2012matrix}, these models are a natural generalization of the multivariate-multiple regression (see also \citealp{anderlucci2014matrix}).
In the mixture framework, finite mixtures of matrix-variate regressions (MN-FMR) have been recently considered by \citet{melnykov2019studying}.
There are no matrix-variate CWMs in the literature and this paper aims to fill this gap by introducing and discussing a matrix-variate CWM in which the cluster-specific marginal distribution of the covariates, and the cluster-specific conditional distribution of the responses given the covariates, are assumed to be matrix normal.

The remainder of the paper is organized as follows.
In Section~\ref{sec:back}, some preliminary aspects are described.
The matrix normal cluster-weighted model (MN-CWM) and the expectation-conditional maximization (ECM) algorithm for parameter estimation are discussed in Section~\ref{sec:model}. 
Computational and operational aspects are laid out in Section~\ref{sec:comp}.
In the simulation study outlined in Section~\ref{sec:artif}, the parameter recovery and the classification performance of the MN-CWM are investigated as well as the capability of the Bayesian information criterion (BIC; \citealp{schwarz1978estimating}) to detect the underlying group structure.
The MN-CWM is also therein compared to the MN-FMR and to the multivariate-multiple normal CWM (MMN-CWM).
The application of the MN-CWM to two real datasets concerning educational indicators and the Italian non-life insurance market is therefore analyzed in Section~\ref{sec:data}, whereas some conclusions and ideas for future developments are drawn in Section~\ref{sec:conclude}.

\section{Background}
\label{sec:back}

\subsection{Matrix normal distribution}
\label{subsec:MXVN}

A $p \times r$ continuous random matrix $\bY \in \real^{p \times r}$ has a matrix normal (MN) distribution, denoted by $\pazocal{N}_{p \times r} \left(\bM,\bPhi,\bPsi\right)$, if its density is 
\begin{equation}
\resizebox{.94\textwidth}{!}{$%
\bphi_{p \times r}\left(\bY;\bM,\bPhi,\bPsi\right)=\left(2\pi\right)^{-\frac{pr}{2}}\left|\bPhi\right|^{-\frac{r}{2}}\left|\bPsi\right|^{-\frac{p}{2}}\exp\left\{-\frac{1}{2}\tr\left[\bPhi^{-1}\left(\bY-\bM\right)\bPsi^{-1}\left(\bY-\bM\right)^\top\right]\right\},
$}
\label{eq:MatNorm}
\end{equation}
where $\bM$ is the $p\times r$ mean matrix, and $\bPhi$ and $\bPsi$ are the $p \times p$ and $r \times r$ covariance matrices associated with the $p$ rows and $r$ columns, respectively.
An equivalent definition specifies the $\left(p\times r\right)$-matrix normal distribution as a special case of the $pr$-multivariate normal distribution.
Specifically,
\begin{equation}
\bY \sim \pazocal{N}_{p \times r} \left(\bM,\bPhi,\bPsi\right)\Leftrightarrow \vect\left(\bY\right) \sim \pazocal{N}_{pr}\left(\vect\left(\bM\right),\bPsi \otimes \bPhi \right),
\label{eq:vectorize}
\end{equation}
where $\pazocal{N}_{pr}\left(\cdot\right)$ denotes the $pr$-variate normal distribution, $\vect\left(\cdot\right)$ is the vectorization operator and $\otimes$ denotes the Kronecker product.
However, a MN distribution has the desirable feature of simultaneously modeling and identify the between and within-variables variabilities as well as reducing the number of free parameters from $pr\left(pr+1\right)/2$ to $\left[r\left(r+1\right)/2\right]+\left[p\left(p+1\right)/2\right]-1$.

\subsection{The matrix-variate regression model}
\label{subsec:MatReg}

Let $\bY \in \real^{p \times r}$ be a continuous random matrix of dimension $p \times r$, containing $p$ responses measured in $r$ occasions.
Suppose we observe a set of $q$ covariates for each occasion, inserted in a matrix $\bX$ of dimension $q \times r$.
A generic matrix-variate regression model for $\bY$ has the form
\begin{equation}
\bY = \bbeta\bw^{\top} + \bB \bX + \bU ,
\label{eq:regModel1}
\end{equation}
where $\bbeta$ is the $p \times 1 $ vector consisting in the parameters related with the intercept, $\bw$ is a $r \times 1$ vector of ones, $\bB$ is the $p \times q$ matrix containing the parameters related to the $q$ covariates and $\bU$
is the $p \times r$ error term matrix.
Model~\eqref{eq:regModel1} can be expressed in compact notation as 
\begin{equation}
\bY = \bB^{*} \bX^{*} + \bU ,
\label{eq:regModel2}
\end{equation}
where $\bB^{*}$ is the $p \times \left(1+q\right)$ matrix involving all the parameters to be estimated and $\bX^{*}$ is the $\left(1+q\right) \times r$ matrix containing the information about the intercept and $q$ covariates \citep{viroli2012matrix}.
If we assume $\bU \sim \pazocal{N}_{p \times r}\left(\b0,\bPhi,\bPsi\right)$, then $\bY|\bX^{*} \sim \pazocal{N}_{p \times r}\left(\bB^{*} \bX^{*},\bPhi,\bPsi\right)$.
Therefore, a matrix-variate regression can be viewed as an encompassing framework containing as special cases the multivariate-multiple regression, when $r = 1$, and the univariate-multiple regression when $r = 1$ and $p = 1$.

\section{Methodology}
\label{sec:model}

\subsection{The matrix normal CWM}
\label{subsec:MatCWM}

Let $\left(\bX,\bY\right)$ be a pair of random matrices, defined as in Section~\ref{subsec:MatReg}, with joint distribution $p\left(\bX,\bY\right)$.
Then, a general matrix CWM has the following joint distribution
\begin{equation}
p\left(\bX,\bY\right) = \sum\limits_{g=1}^{G} p_{g}\left(\bY|\bX^{*}\right) p_{g}\left(\bX\right) \pi_{g},
\label{eq:CWM}
\end{equation}
where $p_{g}\left(\bY|\bX^{*}\right)$ is the cluster-specific conditional distribution of the responses, $p_{g}\left(\bX\right)$ is the cluster-specific marginal distribution of the covariates, and $\pi_{g}$ is the mixing weight (with $\pi_{g} > 0 $ and $\sum\limits_{g=1}^{G} \pi_{g} =1$).
Furthermore, we assume that in each group the conditional expectation $\E\left(\bY|\bX^{*}\right)$ is a linear function of $\bX^{*}$ depending on some parameters. 

In this paper, we focus on model~\eqref{eq:CWM} by assuming that both $p_{g}\left(\bY|\bX^{*}\right)$ and $p_{g}\left(\bX\right)$ are matrix normal densities, and $\E\left(\bY|\bX^{*}\right) = \bB^{*} \bX^{*}$, as described in Section~\ref{subsec:MatReg}.
Thus, model~\eqref{eq:CWM} can be rewritten as
\begin{equation}
\resizebox{.94\textwidth}{!}{$%
p\left(\bX,\bY;\bTheta\right) = \sum\limits_{g=1}^{G} \bphi_{p \times r}\left(\bY|\bX^{*};\bB_{g}^{*}\;\bX^{*},\bPhi_{\bY_{g}},\bPsi_{\bY_{g}}\right) \bphi_{q \times r}\left(\bX;\bM_{g},\bPhi_{\bX_{g}},\bPsi_{\bX_{g}}\right) \pi_{g},
$}
\label{eq:NOR_CWM}
\end{equation}
where $\bTheta$ denotes the set of all parameters.
For ease of understanding, a subscript with the variable name is added to the respective covariance matrices.
Notice that there is an identifiability issue concerning the covariance matrices.
Indeed, for a MN distribution, $\bPsi \otimes \bPhi = \bPsi^{*} \otimes \bPhi^{*}$ if $\bPhi^{*}= a\bPhi$ and $\bPsi^{*}= a^{-1}\bPsi$.
As a result, matrices $\bPhi$ and $\bPsi$ are identifiable up to a multiplicative constant $a$ \citep{melnykov2019studying}.
According to \citet{gallaugher17,gallaugher2018finite}, a way to obtain a unique solution is to fix the first diagonal element of the row covariance matrix to 1.
Therefore, in model~\eqref{eq:NOR_CWM} we adopt this approach by setting the first diagonal element of $\bPhi_{\bY_{g}}$ and $\bPhi_{\bX_{g}}$ to $1$.

If the MN-CWM was not available, a possible approach would be to vectorize the matrices and consider the MMN-CWM, of which it is a special case; see Equation~\eqref{eq:vectorize}.
However, such a procedure leads to two main concerns.
The first one is the overparameterization of the vectorized model.
Secondly, this increased number of parameters may have direct effects on model selection, as will be better analyzed in Section~\ref{subsec:simu4}.

\subsection{Parameter estimation}
\label{sec:EM}

Parameter estimation is carried out via the expectation-conditional maximization (ECM) algorithm \citep{meng1997algorithm}.
The only difference with respect to the expectation-maximization (EM) algorithm \citep{dempster1977maximum} is that the M-step is replaced by a sequence of simpler and computationally convenient CM-steps.
The EM algorithm cannot be directly implemented since there is no closed form solution for the covariance matrices of the MN distribution, i.e.~one of the two depends on the value of the other at the previous iteration \citep{dutilleul1999mle}.

Let $\Sa = \left\{\left(\bX_{i},\bY_{i}\right)\right\}_{i=1}^{N}$ be a sample of $N$ independent observations from model~\eqref{eq:NOR_CWM}.
Then, the incomplete-data likelihood function is
\begin{equation}
\resizebox{.94\textwidth}{!}{$%
  \begin{aligned}
	L\left(\bTheta|\Sa\right) = & \prod_{i=1}^{N} p\left(\bX_{i},\bY_{i};\bTheta\right)\\
                            = & \prod_{i=1}^{N}\left[\sum\limits_{g=1}^{G} \bphi_{p \times r}\left(\bY_{i}|\bX_{i}^{*};\bB_{g}^{*}\;\bX_{i}^{*},\bPhi_{\bY_{g}},\bPsi_{\bY_{g}}\right) \bphi_{q \times r}\left(\bX_{i};\bM_{g},\bPhi_{\bX_{g}},\bPsi_{\bX_{g}}\right) \pi_{g}\right].
\end{aligned}$%
}										
\label{eq:lik}
\end{equation}
Within the formulation of mixture models, $\Sa$ is viewed as being incomplete because, for each observation, we do not know its component membership.
Let $\bz_{i} = \left( z_{i1},\ldots,z_{iG}\right)^{\top}$ be the component membership vector such that $z_{ig}= 1$ if $\left(\bX_{i},\bY_{i}\right)$ comes from group $g$ and $z_{ig}= 0$ otherwise.
Now, the complete-data are $\Sa_{c} = \left\{\left(\bX_{i},\bY_{i},\bz_{i}\right)\right\}_{i=1}^{N}$, and the complete-data likelihood is
\begin{equation}
\resizebox{.94\textwidth}{!}{$%
\begin{aligned}
L_{c}\left(\bTheta|\Sa_{c}\right) = \prod_{i=1}^{N} \prod_{g=1}^{G} \left[\bphi_{p \times r}\left(\bY_{i}|\bX_{i}^{*}; \bB_{g}^{*}\;\bX_{i}^{*},\bPhi_{\bY_{g}},\bPsi_{\bY_{g}}\right) \bphi_{q \times r}\left(\bX_{i};\bM_{g},\bPhi_{\bX_{g}},\bPsi_{\bX_{g}}\right) \pi_{g}\right]^{z_{ig}}.
\end{aligned}$%
}
\label{eq:likcomp}
\end{equation}
Therefore, the corresponding complete-data log-likelihood can be written as
%\begin{equation}
%\resizebox{.94\textwidth}{!}{$%
\begin{align}
l_{c}\left(\bTheta|\Sa_{c}\right) = & \sum\limits_{i=1}^{N} \sum\limits_{g=1}^{G} z_{ig} \ln\left(\pi_{g}\right) + \sum\limits_{i=1}^{N} \sum\limits_{g=1}^{G} z_{ig} \ln\left[\bphi_{p \times r}\left(\bY_{i}|\bX_{i}^{*};\bB_{g}^{*}\;\bX_{i}^{*},\bPhi_{\bY_{g}},\bPsi_{\bY_{g}}\right)\right]\nonumber\\ 
& + \sum\limits_{i=1}^{N} \sum\limits_{g=1}^{G} z_{ig} \ln\left[\bphi_{q \times r}\left(\bX_{i};\bM_{g},\bPhi_{\bX_{g}},\bPsi_{\bX_{g}}\right)\right].\label{eq:llk}
\end{align}%
%}							
%\end{equation}
In the following, by adopting the notation used in \citet{tomarchio2020two}, the parameters marked with one dot correspond to the updates at the previous iteration and those marked with
two dots represent the updates at the current iteration.

\paragraph{E-Step} 
\label{subsec:E-step}

The E-step requires calculation of the conditional expectation of \eqref{eq:llk}, given the observed data and the current estimate of the parameters $\dot{\bTheta}$.
To do this, we need to calculate 
%\begin{equation}
%\resizebox{.925\textwidth}{!}{$%
\begin{align}
\ddot{z}_{ig} & = \E_{\dot{\bTheta}} \left[Z_{ig}|\bX_{i},\bY_{i}\right]\nonumber\\
                          &= \frac{\dot{\pi}_{g} \bphi_{p \times r}(\bY_{i}|\bX_{i}^{*};\dot{\bB}_{g}^{*}\bX_{i}^{*},\dot{\bPhi}_{\bY_{g}},\dot{\bPsi}_{\bY_{g}}) \bphi_{q \times r}(\bX_{i};\dot{\bM}_{g},\dot{\bPhi}_{\bX_{g}},\dot{\bPsi}_{\bX_{g}})}  {\sum\limits_{j=1}^{G} \dot{\pi}_{j} \bphi_{p \times r}(\bY_{i}|\bX_{i}^{*};\dot{\bB}_{j}^{*}\bX_{i}^{*},\dot{\bPhi}_{\bY_{j}},\dot{\bPsi}_{\bY_{j}}) \bphi_{q \times r}(\bX_{i};\dot{\bM}_{j},\dot{\bPhi}_{\bX_{j}},\dot{\bPsi}_{\bX_{j}})}, \label{eq:E-step}
\end{align}
%$%
%}						
%\end{equation}
which corresponds to the posterior probability that the unlabeled observation $\left(\bX_{i},\bY_{i}\right)$ belongs to the $g$th component of the mixture.

\paragraph{CM-Step 1}
\label{subsec:cm1}

At the first CM-step, we maximize the expectation of the complete-data log-likelihood with respect to $\bTheta_{1}=\left\{\pi_g,\bM_{g},\bPhi_{\bX_{g}},\bB_{g}^{*},\bPhi_{\bY_{g}}\right\}_{g=1}^{G}$, fixing $\bTheta_{2}=\left\{\bPsi_{\bX_{g}},\bPsi_{\bY_{g}}\right\}_{g=1}^{G}$ at $\dot\bTheta_{2}$.
In particular, we obtain
\begin{equation}
\ddot{\pi}_g   = \frac{\sum_{i=1}^N \ddot{z}_{ig}}{N}, 
\label{eq:cwm_cm1.1} 
\end{equation}
\begin{equation}
\ddot{\bM}_g    = \frac{\sum_{i=1}^N \ddot{z}_{ig}\bX_i}{\sum_{i=1}^N \ddot{z}_{ig}}, 
\label{eq:cwm_cm1.2} 
\end{equation}
\vspace{-8mm}
\begin{align}
\ddot{\bB}_{g}^{*} & = \left[\sum_{i=1}^N \ddot{z}_{ig} \bY_{i} (\dot{\bPsi}_{\bY_{g}})^{-1} \bX_{i}^{*\top}\right] \left[\sum_{i=1}^N \ddot{z}_{ig} \bX_{i}^{*} (\dot{\bPsi}_{\bY_{g}})^{-1} \bX_{i}^{*\top}\right]^{-1}, \label{eq:cwm_cm1.3} \\
\ddot{\bPhi}_{\bX_{g}} & = \frac{\sum_{i=1}^N \ddot{z}_{ig}\left(\bX_i-\ddot{\bM}_g\right)\left(\dot\bPsi_{\bX_{g}}\right)^{-1}\left(\bX_i-\ddot{\bM}_g\right)^{\top}}{r \sum_{i=1}^N \ddot{z}_{ig}}, \label{eq:cwm_cm1.4} \\
\ddot{\bPhi}_{\bY_{g}} & = \frac{\sum_{i=1}^N \ddot{z}_{ig}\left(\bY_i-\ddot{\bB}^{*}_{g}\bX_{i}^{*}\right)\left(\dot\bPsi_{\bY_{g}}\right)^{-1}\left(\bY_i-\ddot{\bB}^{*}_{g}\bX_{i}^{*}\right)^{\top}}{r \sum_{i=1}^N \ddot{z}_{ig}}. \label{eq:cwm_cm1.5} 
\end{align}

\paragraph{CM-Step 2}

At the second CM-step, we maximize the expectation of the complete-data log-likelihood with respect to $\bTheta_{2}$, keeping fixed $\bTheta_{1}$ at $\ddot\bTheta_{1}$.
Therefore, we have
\begin{align}
\ddot{\bPsi}_{\bX_{g}} & = \frac{\sum_{i=1}^N \ddot{z}_{ig}\left(\bX_i-\ddot{\bM}_g\right)^{\top}\left(\ddot\bPhi_{\bX_{g}}\right)^{-1}\left(\bX_i-\ddot{\bM}_g\right)}{q \sum_{i=1}^N \ddot{z}_{ig}},\label{eq:cwm_cm2.1} \\
\ddot{\bPsi}_{\bY_{g}} & = \frac{\sum_{i=1}^N \ddot{z}_{ig}\left(\bY_i-\ddot{\bB}^{*}_{g}\bX_{i}^{*}\right)^{\top}\left(\ddot\bPhi_{\bY_{g}}\right)^{-1}\left(\bY_i-\ddot{\bB}^{*}_{g}\bX_{i}^{*}\right)}{p \sum_{i=1}^N \ddot{z}_{ig}}.\label{eq:cwm_cm2.2} 
\end{align}

\section{Computational and operational aspects}
\label{sec:comp}

The code for the ECM algorithm previously described, and for the analysis in the following sections, is written within the R computing environment \citep{R_software}.

\subsection{ECM initialization}
\label{subsec:init}

The choice of the starting values is an important aspect for EM-based algorithms (see, e.g., \citealp{maitra2010simulating, michael2016effective}).
A common way to start an EM-based algorithm consists in providing initial values for the $z_{ig}$ in~\eqref{eq:E-step}, during the first E-step of the algorithm \citep{mclachlan2000finite}.
In our case, we need to provide an initial value also for $\bPsi_{\bX_{g}}$ and $\bPsi_{\bY_{g}}$ in~\eqref{eq:cwm_cm1.4} and~\eqref{eq:cwm_cm1.5}, respectively, during the first CM-step 1 of the algorithm.
Therefore, the following initialization strategy is implemented:
\begin{enumerate}
\item Generate $G$ random positive definite matrices for both $\bPsi_{\bX_{g}}$ and $\bPsi_{\bY_{g}}$.
This is done via the \texttt{genPositiveDefMat()} function of the \textbf{clusterGeneration} package, by using the ``eigen'' method. For further details see \citet{clusterG}. 
\item generate $N$ random vectors $z_{i}=\left(z_{i1},\ldots,z_{iG}\right)^{\top}$, $i=1,\ldots,N$.
This is done by using the following three approaches:
\begin{enumerate}[label*=\arabic*.]
    \item in a ``soft'' way, by generating $G$ positive random values from a uniform distribution on [0,1] for each observation, that are subsequently normalized to have a unitary sum.
		Being purely random, this procedure is repeated 15 times, and the solution maximizing the observed-data log-likelihood among these runs is considered;
		\item in a ``hard'' way, by using the classification produced by the $k$-means algorithm on the vectorized and merged data. 
		Specifically, after computing $\left\{\vect\left(\bX_{i}\right)\right\}_{i=1}^{N}$ and $\left\{\vect\left(\bY_{i}\right)\right\}_{i=1}^{N}$, the data are merged so that for each 
		observation we have a vector of dimension $\left(pr+qr\right) \times 1$;
		\item in a ``hard'' way, by using the classification produced by mixtures of matrix-normal distributions, computed on the merged data.
		In detail, for each observation we have a $\left(p+q\right) \times r$ matrix.
  \end{enumerate}
\end{enumerate}
The approach producing the highest observed-data log-likelihood is finally selected.

\subsection{Spurious clusters}
\label{subsec:spur}

A well-known issue in mixture models is related to the possibility for EM-based algorithms to converge to spurious local maximizers.
This is not a failing of these algorithms, and such solutions reflect a random pattern in the data rather than an underlying group structure.
They can be usually detected by the presence of data groups with very few observations or small eigenvalues compared to the other ones.
Consequently, these solutions often have a high likelihood, but are of little practical use or real-world interpretation \citep{mclachlan2000finite}.
Various approaches for mitigate this problem have been proposed in the literature.
For example, \citet{leisch2004flexmix} and \citet{dang2015families} impose a minimum size for the clusters of $\pi_{g}=0.05$.
\citet{melnykov2019studying} fought with this issue by excluding all the solutions that involved clusters consisting of less than 5 points. 
We faced this problem by removing the solutions that included clusters with estimated ${\pi}_{g}\approx 0.05$ or less, and with eigenvalues close to zero, in one or more of its estimated covariance matrices.

\subsection{Model selection and performance evaluation}
\label{subsec:classperf}

It is often the case that the number of groups $G$ is not known in advance, and model selection criteria are commonly used for estimating it.
Among them, the BIC is one of the most popular, and will be used in the following.
Furthermore, when the true classification of the data is known, the adjusted Rand index (ARI; \citealp{hubert1985comparing}) can be adopted to evaluate the clustering performance of a model.
Specifically, the ARI calculates the agreement between the true classification and the one predicted by the model.
An ARI of 1 indicates perfect agreement between the two partitions, whereas the expected value of the ARI under a random classification is 0.
It will be used in the manuscript along with the misclassification percentage rate $\eta$, which is the percentage of units classified in the wrong classes.

\section{Simulation studies}
\label{sec:artif}

\subsection{Simulation 1: A focus on the matrix-normal CWM}
\label{subsec:simu1}

In this study, several aspects related to our model are analyzed.
First of all, since the ECM algorithm is used to fit the model, it is desirable to evaluate its parameter recovery, i.e.~whether it can recover the generating parameters accurately. 
For this reason, data are generated from a four-component MN-CWM with $p=q=r=3$.
Two scenarios are then evaluated, according to the different level of overlap of the mixture components.
In the first scenario (labeled as ``Scenario A$_{1}$''), the mixture components are well-separated both in $\bX$, by assuming relatively distant mean matrices, and in $\bY|\bX^{*}$, by using different intercepts and slopes.
On the contrary, in the second scenario (labeled as ``Scenario B$_{1}$''), there is a certain amount of overlap, since the intercepts are all equal among the mixture components, while the slopes and the mean matrices assume approximately the same values among the mixture components.
The parameters used for Scenario A$_{1}$ are displayed in~\appendixname~\ref{sec:app1}.
Under Scenario B$_{1}$, the set of parameters $\left\{\pi_g, \bPhi_{\bX_{g}}, \bPsi_{\bX_{g}}, \bPhi_{\bY_{g}}, \bPsi_{\bY_{g}}\right\}_{g=1}^{4}$, $\bM_{1}$ and the slopes in $\bB^{*}_{1}$ and $\bB^{*}_{3}$, are the same of Scenario A$_{1}$.
The other mean matrices are obtained by adding a constant $c$ to each element of the corresponding mean matrices used for Scenario A$_{1}$.
In detail, $c$ is equal to -5, 5 and -10 for $\bM_{2}$, $\bM_{3}$ and $\bM_{4}$, respectively.
The intercept column of all the mixture components is equal to $\left(7,2,5\right)^{\top}$, whereas the slopes in $\bB^{*}_{2}$ and $\bB^{*}_{4}$ are all multiplied by -1, with respect to those used in Scenario A$_{1}$.
Lastly, within each scenario two sample sizes are considered, i.e.~$N=200$ and $N=500$.

On all the generated datasets, the MN-CWM is fitted directly with $G=4$, and the bias and mean squared error (MSE) of the parameter estimates are computed.
For brevity's sake, and as also supported by the existing CWM literature (see, e.g.~\citealp{punzo2014flexible,ingrassia2015generalized,punzo2016clustering,punzo2017robust}), the attention will be only focused on the regression coefficients $\left\{\bB_{g}^{*}\right\}_{g=1}^{G}$.
Before showing the obtained results, it is important to underline the well-known label switching issue, caused by the invariance of a mixture distribution to relabeling the components \citep{fruhwirth2006finite}.
There are no generally accepted labeling methods.
Then, to assign the correct labels, an analysis of the overall estimated parameters is conducted on each generated dataset to properly identify each mixture component.

\tablename~\ref{tab:resA} summarizes the estimated bias and MSE of the parameter estimates for Scenario A$_{1}$, over one hundred replications for each sample size $N$, after fitting the MN-CWM with $G=4$.
The same is reported for Scenario B$_{1}$ in \tablename~\ref{tab:resB}.
The first and most immediate result is that the biases and the MSEs assume very small values in both scenarios.
This is particularly relevant for Scenario B$_{1}$, because of the presence of overlap between the mixture components.
Furthermore, within in each scenario, an increase in the sample size leads to a rough improvement of the parameter estimates, whereas it systematically reduces the MSE.
\begin{table}[!ht]
\caption{Estimated bias and MSE of the regression coefficients $\left\{\bB_{g}^{*}\right\}_{g=1}^{4}$, over 100 replications, under Scenario A$_{1}$.} 
\centering
\resizebox{0.75\columnwidth}{!}{%  
\begin{tabular}{cccc} 
\toprule
&&$N=200$&$N=500$\\ 
\midrule 
\addlinespace[3mm]
\multirow{4}{*}{Group 1} & Bias & $\setlength\arraycolsep{2.8pt}\begin{pmatrix*}[r]  0.032 & -0.001 & 0.005 & 0.002 \\
                                                                                    -0.025 & -0.010 &-0.008 &-0.004 \\
																																										-0.028 &  0.006 &-0.014 &-0.004 \end{pmatrix*}$ 
																& $\setlength\arraycolsep{2.8pt}\begin{pmatrix*}[r]  0.001 & -0.003 &-0.002 & 0.007 \\
                                                                                    -0.033 &  0.002 &-0.007 & 0.004 \\
																																										 0.003 & -0.004 &-0.002 & 0.005 \end{pmatrix*}$ \\[7mm]
                         & MSE  & $\setlength\arraycolsep{2.8pt}\begin{pmatrix*}[r]  0.343 & 0.011 & 0.018 & 0.014 \\
                                                                                     0.337 & 0.010 & 0.017 & 0.018 \\
																																										 0.365 & 0.011 & 0.020 & 0.016 \end{pmatrix*}$ 
																& $\setlength\arraycolsep{2.8pt}\begin{pmatrix*}[r]  0.111 & 0.004 & 0.006 & 0.007 \\
                                                                                     0.114 & 0.004 & 0.006 & 0.006 \\
																																										 0.104 & 0.004 & 0.006 & 0.005 \end{pmatrix*}$ \\[10mm]
\midrule
\addlinespace[3mm]
\multirow{4}{*}{Group 2} & Bias & $\setlength\arraycolsep{2.8pt}\begin{pmatrix*}[r]  0.039 & -0.004 & 0.001 &-0.001 \\
                                                                                    -0.005 & -0.004 & 0.002 & 0.006 \\
																																										-0.004 & -0.008 &-0.004 & 0.009 \end{pmatrix*}$ 
																& $\setlength\arraycolsep{2.8pt}\begin{pmatrix*}[r]  0.001 & -0.003 &-0.002 &-0.001 \\
                                                                                     0.019 & -0.000 &-0.000 &-0.000 \\
																																										 0.042 & -0.004 &-0.002 &-0.001 \end{pmatrix*}$ \\[7mm]
                         & MSE  & $\setlength\arraycolsep{2.8pt}\begin{pmatrix*}[r]  0.252 & 0.003 & 0.003 & 0.004 \\
                                                                                     0.204 & 0.002 & 0.003 & 0.004 \\
																																										 0.170 & 0.002 & 0.003 & 0.005 \end{pmatrix*}$ 
																& $\setlength\arraycolsep{2.8pt}\begin{pmatrix*}[r]  0.084 & 0.001 & 0.001 & 0.001 \\
                                                                                     0.089 & 0.001 & 0.001 & 0.001 \\
																																										 0.099 & 0.001 & 0.001 & 0.001 \end{pmatrix*}$ \\[10mm]
\midrule
\addlinespace[3mm]
\multirow{4}{*}{Group 3} & Bias & $\setlength\arraycolsep{2.8pt}\begin{pmatrix*}[r]  0.005 &-0.008 & 0.005 & 0.001 \\
                                                                                    -0.020 &-0.010 &-0.002 &-0.000 \\
																																										-0.051 &-0.008 &-0.003 &-0.001 \end{pmatrix*}$ 
																& $\setlength\arraycolsep{2.8pt}\begin{pmatrix*}[r]  0.002 &-0.000 &-0.001 &-0.000 \\
                                                                                     0.055 & 0.001 & 0.004 & 0.001 \\
																																										 0.018 & 0.003 & 0.002 &-0.001 \end{pmatrix*}$ \\[7mm]
                         & MSE  & $\setlength\arraycolsep{2.8pt}\begin{pmatrix*}[r]  0.229 & 0.005 & 0.002 & 0.003 \\
                                                                                     0.244 & 0.005 & 0.002 & 0.003 \\ 
																																										 0.235 & 0.006 & 0.002 & 0.004 \end{pmatrix*}$ 
																& $\setlength\arraycolsep{2.8pt}\begin{pmatrix*}[r]  0.104 & 0.002 & 0.001 & 0.001 \\
                                                                                     0.122 & 0.002 & 0.001 & 0.001 \\ 
																																										 0.111 & 0.002 & 0.001 & 0.001 \end{pmatrix*}$ \\[10mm]
\midrule
\addlinespace[3mm]
\multirow{4}{*}{Group 4} & Bias & $\setlength\arraycolsep{2.8pt}\begin{pmatrix*}[r]  0.097 &-0.008 & 0.011 &-0.005 \\
                                                                                     0.027 &-0.006 & 0.006 & 0.002 \\
																																										-0.017 &-0.005 & 0.006 & 0.005 \end{pmatrix*}$ 
																& $\setlength\arraycolsep{2.8pt}\begin{pmatrix*}[r] -0.041 & 0.003 & 0.001 &-0.002 \\
                                                                                    -0.045 & 0.003 & 0.005 &-0.002 \\
																																										-0.006 & 0.002 & 0.003 &-0.004 \end{pmatrix*}$ \\[7mm]
                         & MSE  & $\setlength\arraycolsep{2.8pt}\begin{pmatrix*}[r]  0.412 & 0.003 & 0.005 & 0.004 \\
                                                                                     0.412 & 0.003 & 0.007 & 0.004 \\
																																										 0.397 & 0.003 & 0.006 & 0.004 \end{pmatrix*}$ 
																& $\setlength\arraycolsep{2.8pt}\begin{pmatrix*}[r]  0.242 & 0.001 & 0.001 & 0.001 \\
                                                                                     0.200 & 0.001 & 0.001 & 0.001 \\
																																										 0.209 & 0.001 & 0.002 & 0.001 \end{pmatrix*}$ \\[7mm]
\bottomrule 
\end{tabular}%
}
\label{tab:resA} 
\end{table}
\begin{table}[!ht]
\caption{Estimated bias and MSE of the regression coefficients $\left\{\bB_{g}^{*}\right\}_{g=1}^{G}$, over 100 replications, under Scenario B$_{1}$.} 
\centering
\resizebox{0.75\columnwidth}{!}{% 
\begin{tabular}{cccc} 
\toprule
&&$N=200$&$N=500$\\ 
\midrule 
\addlinespace[3mm]
\multirow{4}{*}{Group 1} & Bias & $\setlength\arraycolsep{2.8pt}\begin{pmatrix*}[r] -0.058 & -0.011 &-0.015 & 0.015 \\
                                                                                    -0.037 & -0.008 &-0.011 & 0.018 \\
																																										-0.082 & -0.002 &-0.027 & 0.010 \end{pmatrix*}$ 
																& $\setlength\arraycolsep{2.8pt}\begin{pmatrix*}[r]  0.052 & -0.001 & 0.008 &-0.002 \\
                                                                                     0.034 & -0.001 & 0.004 & 0.002 \\
																																										-0.018 &  0.000 &-0.006 & 0.001 \end{pmatrix*}$ \\[7mm]
                         & MSE  & $\setlength\arraycolsep{2.8pt}\begin{pmatrix*}[r]  0.368 & 0.014 & 0.018 & 0.021 \\
                                                                                     0.410 & 0.014 & 0.021 & 0.022 \\
																																										 0.361 & 0.011 & 0.019 & 0.022 \end{pmatrix*}$ 
																& $\setlength\arraycolsep{2.8pt}\begin{pmatrix*}[r]  0.118 & 0.004 & 0.007 & 0.006 \\
                                                                                     0.117 & 0.004 & 0.007 & 0.007 \\
																																										 0.124 & 0.004 & 0.006 & 0.007 \end{pmatrix*}$ \\[10mm]
\midrule
\addlinespace[3mm]
\multirow{4}{*}{Group 2} & Bias & $\setlength\arraycolsep{2.8pt}\begin{pmatrix*}[r] -0.046 &  0.002 &-0.013 &-0.001 \\
                                                                                    -0.037 &  0.005 & 0.001 & 0.001 \\
																																										-0.013 &  0.008 & 0.004 & 0.005 \end{pmatrix*}$ 
																& $\setlength\arraycolsep{2.8pt}\begin{pmatrix*}[r] -0.014 &  0.006 &-0.002 & 0.005 \\
                                                                                    -0.030 &  0.004 &-0.006 & 0.008 \\
																																										-0.008 &  0.000 &-0.002 & 0.009 \end{pmatrix*}$ \\[7mm]
                         & MSE  & $\setlength\arraycolsep{2.8pt}\begin{pmatrix*}[r]  0.046 & 0.003 & 0.006 & 0.004 \\
                                                                                     0.046 & 0.004 & 0.003 & 0.005 \\
																																										 0.043 & 0.004 & 0.004 & 0.005 \end{pmatrix*}$ 
																& $\setlength\arraycolsep{2.8pt}\begin{pmatrix*}[r]  0.015 & 0.001 & 0.001 & 0.002 \\
                                                                                     0.013 & 0.001 & 0.001 & 0.001 \\
																																										 0.013 & 0.001 & 0.001 & 0.002 \end{pmatrix*}$ \\[10mm]
\midrule
\addlinespace[3mm]
\multirow{4}{*}{Group 3} & Bias & $\setlength\arraycolsep{2.8pt}\begin{pmatrix*}[r]  0.035 &-0.017 & 0.011 & 0.016 \\
                                                                                     0.011 &-0.006 & 0.012 & 0.008 \\
																																										 0.023 &-0.010 & 0.005 & 0.010 \end{pmatrix*}$ 
																& $\setlength\arraycolsep{2.8pt}\begin{pmatrix*}[r]  0.007 &-0.004 & 0.002 &-0.000 \\
                                                                                     0.015 &-0.004 & 0.001 & 0.000 \\
																																										 0.028 &-0.005 & 0.004 &-0.000 \end{pmatrix*}$ \\[7mm]
                         & MSE  & $\setlength\arraycolsep{2.8pt}\begin{pmatrix*}[r]  0.078 & 0.006 & 0.003 & 0.003 \\
                                                                                     0.073 & 0.005 & 0.003 & 0.003 \\ 
																																										 0.080 & 0.005 & 0.002 & 0.003 \end{pmatrix*}$ 
																& $\setlength\arraycolsep{2.8pt}\begin{pmatrix*}[r]  0.027 & 0.002 & 0.001 & 0.001 \\
                                                                                     0.025 & 0.002 & 0.001 & 0.001 \\ 
																																										 0.030 & 0.002 & 0.001 & 0.001 \end{pmatrix*}$ \\[10mm]
\midrule
\addlinespace[3mm]
\multirow{4}{*}{Group 4} & Bias & $\setlength\arraycolsep{2.8pt}\begin{pmatrix*}[r]  0.039 & 0.002 & 0.002 &-0.003 \\
                                                                                     0.005 &-0.001 &-0.004 &-0.007 \\
																																										-0.051 &-0.003 & 0.004 & 0.004 \end{pmatrix*}$ 
																& $\setlength\arraycolsep{2.8pt}\begin{pmatrix*}[r] -0.043 & 0.004 & 0.001 &-0.003 \\
                                                                                    -0.014 &-0.000 & 0.003 & 0.002 \\
																																										-0.008 &-0.002 & 0.002 & 0.003 \end{pmatrix*}$ \\[7mm]
                         & MSE  & $\setlength\arraycolsep{2.8pt}\begin{pmatrix*}[r]  0.147 & 0.003 & 0.005 & 0.004 \\
                                                                                     0.160 & 0.006 & 0.007 & 0.006 \\
																																										 0.132 & 0.003 & 0.007 & 0.006 \end{pmatrix*}$ 
																& $\setlength\arraycolsep{2.8pt}\begin{pmatrix*}[r]  0.061 & 0.001 & 0.002 & 0.002 \\
                                                                                     0.060 & 0.001 & 0.002 & 0.002 \\
																																										 0.069 & 0.001 & 0.002 & 0.002 \end{pmatrix*}$ \\[7mm]
\bottomrule 
\end{tabular}%
}
\label{tab:resB} 
\end{table}

Other aspects that are investigated consist in the evaluation of the classification produced by our model, as well as the capability of the BIC in identifying the correct number of groups in the data.
For this reason, under each of the considered scenarios, the MN-CWM is fitted to the generated datasets for $G\in\left\{1,2,3,4,5\right\}$, and the results are reported in \tablename~\ref{tab:bic}.
\begin{table}[!ht]
\caption{$\overline{\text{ARI}}$ and $\overline{\eta}$ values, along with the number of times the true $G$ is selected by the BIC, over 100 replications, for scenarios A$_{1}$ and B$_{1}$.} 
\centering 
\begin{tabular}{c|ccc|ccc} 
\toprule
& \multicolumn{3}{c|}{$N=200$} &   \multicolumn{3}{c}{$N=500$} \\
\midrule 
 & $\overline{\text{ARI}}$ & $\overline{\eta}$ & True $G$ & $\overline{\text{ARI}}$ & $\overline{\eta}$ & True $G$ \\ 
\hline 
Scenario A$_{1}$ & 1.00 & 0.00\% & 100 & 1.00 & 0.00\% & 100 \\ 
Scenario B$_{1}$ & 0.91 & 3.04\% & 99  & 0.92 & 2.71\% & 100 \\ 
\bottomrule 
\end{tabular}
\label{tab:bic} 
\end{table}
It is easy to see that under scenario A$_{1}$, a perfect classification is always obtained, regardless of the chosen sample size.
Additionally the BIC regularly detects the correct number of groups in the data.
Under scenario B$_{1}$, because of the larger overlap, the $\overline{\text{ARI}}$ assumes lower but in any case good values.
Relatedly, the percentage of misclassified units stands at around the 3\% for both sample sizes.
About the BIC, also in this case it properly identifies the underlying group structure, with only one exception when $N=200$.

A final aspect that is evaluated in this study concerns the initialization strategy.
Specifically, \tablename~\ref{tab:init} displays the number of times each strategy for the $z_{i}$ produces the highest log-likelihood at convergence, within each scenario and for both sample sizes.
The initial $G$ random matrices for $\bPsi_{\bX_{g}}$ and $\bPsi_{\bY_{g}}$ are assumed to be the same.

\begin{table}[!ht]
\caption{Number of times, over 100 replications, the considered initialization strategies produced the highest log-likelihood at convergence.} 
\centering 
\begin{tabular}{c|ccc|ccc} 
\toprule
& \multicolumn{3}{c|}{$N=200$} &   \multicolumn{3}{c}{$N=500$} \\
\midrule 
 & Random  & $k$-means & Mixture & Random & $k$-means & Mixture \\ 
\hline 
Scenario A$_{1}$ & 100 & 77 & 98 & 100 & 79 & 95 \\ 
Scenario B$_{1}$ & 97 & 74 & 87 & 100 & 83 & 100 \\ 
\bottomrule 
\end{tabular}
\label{tab:init} 
\end{table}
The first result suggests the importance of considering multiple initialization strategies, since none of them are preferred in all the generated datasets.
However, the random strategy is quite close to this target, since it only fails in 3 datasets under scenario B$_{1}$.
Very similar performances are obtained when the mixture initialization is used.
On the contrary, the $k$-means strategy provides the worst performances, even if it produces the best solution in approximately the 80\% of the datasets.

\subsection{Simulation 2: A comparison between the matrix-normal CWM and the matrix-normal FMR}
\label{subsec:simu3}

In this study, the matrix-normal CWM is compared to the matrix-normal FMR.
Specifically, three scenarios with $N=200$, $p=2$, $q=3$ and $r=4$ are considered, and in each of them thirty datasets from a matrix-normal CWM with $G=2$ are generated.
The first scenario (hereafter simply referred to as ``Scenario A$_{2}$'') is characterized by the fact that the two groups differ only for the intercepts and the covariance matrices.
This implies that they have totally overlapped mean matrices, which should make the distribution of the covariates $p_{g}\left(\bX\right)$ not very important for clustering.
The parameters used to generate the datasets are displayed in~\appendixname~\ref{sec:app2}.
In the second scenario (``Scenario B$_{2}$'') the two groups have the same $\bB^{*}_{g}$ and $\pi_g$.
The parameters used to generate the datasets are the same as for Scenario A$_{2}$, but with only two differences: 
a value $c=5$ is added to each element of $\bM_{2}$, and $\bB^{*}_{2}=\bB^{*}_{1}$.
Lastly, in the third scenario (``Scenario C$_{2}$''), the two groups have only the same slopes and $\pi_g$.
Here, with respect to the parameters used under Scenario B$_{2}$, the only difference is in the intercepts vectors which are $\left(-3,-4\right)^{\top}$ and $\left(-7,-8\right)^{\top}$, for the first and the second group, respectively.

The MN-CWM and the MN-FMR are then fitted to the datasets of each scenario for $G\in\left\{1,2,3\right\}$, and the results in terms of model selection and clustering are reported in \tablename~\ref{tab:bic2}.
It is possible to see that in Scenario A$_{2}$ the BIC correctly selects two groups for both models and the classifications produced are perfect.
Therefore, even if the two groups have the same means and are strongly overlapped, the MN-CWM seems able to properly identify the true underlying group structure.
However, under such scenario the MN-FMR should be preferred, since the distribution of the covariates $p_{g}\left(\bX\right)$ is not useful for clustering, and it is more parsimonious than the MN-CWM.
On the contrary, Scenarios B$_{2}$ and C$_{2}$ represent typical examples of the usefulness of $p_{g}\left(\bX\right)$.
Specifically, the BIC always identifies just one group under both scenarios for the MN-FMR, with obvious consequences in terms of the classification produced.
Notice that, even if the MN-FMR had been fitted directly with $G=2$, the resulting classifications would lead to almost identical $\overline{\text{ARI}}$ and $\overline{\eta}$ for Scenario B$_{2}$, and slightly better performance for Scenario C$_{2}$, since $\overline{\text{ARI}}=0.15$ and $\overline{\eta}=32.48\%$. 
This underlines how regardless of the BIC, the MN-FMR is not able to properly model such data structures.
\begin{table}[!ht]
\caption{$\overline{\text{ARI}}$ and $\overline{\eta}$ values, along with the number of times the true $G$ is selected by the BIC, over 30 replications, , for scenarios A$_{2}$, B$_{2}$ and C$_{3}$.
% in each scenario, for the MN-CWM and MN-FMR.
} 
\centering 
\begin{tabular}{c|ccc|ccc} 
\toprule
& \multicolumn{3}{c|}{MN-CWM} & \multicolumn{3}{c}{MN-FMR} \\
\midrule 
 & $\overline{\text{ARI}}$ & $\overline{\eta}$ & True $G$ & $\overline{\text{ARI}}$ & $\overline{\eta}$ & True $G$ \\ 
\hline 
Scenario A$_{2}$ & 1.00 & 0.00\% & 100 & 1.00 & 0.00\%  & 100 \\ 
Scenario B$_{2}$ & 0.99 & 0.03\% & 100 & 0.00 & 47.22\% & 0 \\ 
Scenario C$_{2}$ & 1.00 & 0.01\% & 100 & 0.00 & 47.18\% & 0 \\ 
\bottomrule 
\end{tabular}
\label{tab:bic2} 
\end{table}

\subsection{Simulation 3: A comparison between the matrix-normal CWM and the multivariate-multiple normal CWM}
\label{subsec:simu4}

In this study, the MN-CWM is compared to the MMN-CWM.
To show the effects of data vectorization, we consider two experimental factors: the matrix dimensionality and the number of groups $G$.
About the dimensionality, we assume square matrices having the same dimensions both for the responses and the covariates, i.e.~$p=q=r \in \left\{2,3,4\right\}$.
Similarly, situations with three different number of groups are evaluated, i.e.~$G\in\left\{2,3,4\right\}$.
By combining both experimental factors, 9 scenarios are obtained, and for each of them thirty datasets are generated from a MN-CWM.
The parameters used to generated the data comes from Section~\ref{subsec:simu1} and are shown in~\appendixname~\ref{sec:app1}.
In detail, when $p=q=r=2$ they are obtained by taking the submatrix in the upper-left corner of each parameter, when $p=r=q=3$ they are exactly as displayed, whereas when $p=r=q=4$ a row and a column on each parameter matrix is added, which for brevity's sake are not here reported.
About the number of groups, and by considering \appendixname~\ref{sec:app1}, when $G=2$ and $G=3$ the first two and three groups are selected, respectively, while when $G=4$ all of them are considered.

The MN-CWM is then fitted to each dataset for~$G\in\left\{1,2,3,4,5\right\}$.
The same is done for the MMN-CWM after data vectorization, and the results of both models in terms of model selection by the BIC are shown in \tablename~\ref{tab:vecc}.
As we can see, when the MN-CWM is considered, regardless of the data dimensionality and the number of groups, the BIC always selects the correct number of groups.
The same also holds for the MMN-CWM when $p=q=r=2$ or, regardless of the data dimensionality, when $G=2$.
However, when $p=q=r=3$ the BIC starts to face issues for $G=3$, since the true number of groups is detected only 11 times (the other 19 times $G=2$ is selected), and it systematically fails when $G=4$.
This problem gets even worse when $p=r=q=4$ (with the exclusion of $G=2$).
The reason for such failures is related to the increased number of parameters with respect to the MN-CWM.
Therefore, we have on the one hand a model that can seriously becomes overparametrized, with negative effects also on model selection (the MMN-CWM), and on the other hand a model (the MN-CWM) which is able to fit the same data in a far more parsimonious way and without causing problems on model selection.
\begin{table}[!ht]
\caption{Number of times, over 30 replications, the true $G$ is selected by the BIC in each of the 9 scenarios, for the MN-CWM and MMN-CWM.} 
\centering 
\begin{tabular}{c|ccc|ccc} 
\toprule
& \multicolumn{3}{c|}{MN-CWM} & \multicolumn{3}{c}{MMN-CWM} \\
\midrule 
          & $G=2$ & $G=3$ & $G=4$ & $G=2$ & $G=3$ & $G=4$ \\ 
\hline 
$p=q=r=2$ & 30 & 30 & 30 & 30 & 30 & 30 \\ 
$p=r=q=3$ & 30 & 30 & 30 & 30 & 11 & 0 \\ 
$p=r=q=4$ & 30 & 30 & 30 & 30 & 0  & 0 \\ 
\bottomrule 
\end{tabular}
\label{tab:vecc} 
\end{table}

\section{Real data applications}
\label{sec:data}

\subsection{Education data}
\label{subsec:ANVUR}

The first dataset comes from the Italian national agency for the evaluation of universities and research institutes, which makes available to Italian universities quantitative indicators related to the academic careers of the students and the results of the training activities.
For this application, the following $p=2$ responses, that measure the level of completion of studies by students, are considered
\begin{enumerate}
\item percentage of students that graduate within $T+1$ years (Complete), 
\item percentage of students that drop after $T+1$ years (Drop),
\end{enumerate}
where $T$ is the normal duration of the study program.
Moreover, the following $q=2$ covariates, that may be helpful in explaining this progress,
\begin{enumerate}
\item percentage of course credits earned in the first year over the total to be achieved (Credits), 
\item percentage of students that have earned at least 40 course credits during the solar year (Students),
\end{enumerate}
are taken into account.
For sake of simplicity, hereafter these variables will be referred by using the names given in round brackets.
All the measurements refer to $N=75$ study programs in the non-telematic Italian universities, over $r=3$ years.
Each study program is measured at national level, i.e.~it is the average value of all the study programs of the same type across the country, for the reference period.

There are two groups in the data, namely $N_{1}=33$ bachelor's degrees and $N_{2}=42$ master's degrees.
The MN-CWM and the MN-FMR are fitted to the data for $G\in\left\{1,2,3\right\}$ and their results are reported in \tablename~\ref{tab:Anvur_ari}.
\begin{table}[!htbp]
\caption{Education data: ARI and $\eta$ for the MN-CWM and MN-FMR selected by the BIC.} 
\centering 
\begin{tabular}{c|ccc} 
\toprule
Model & $G$ & ARI & $\eta$ \\ 
\hline 
MN-CWM & 2 & 1.00 &  0.00\% \\ 
MN-FMR & 3 & 0.88 &  6.67\% \\
\bottomrule 
\end{tabular}
\label{tab:Anvur_ari} 
\end{table}
The BIC selects a two-component MN-CWM that yields a perfect classification of the data.
On the contrary, a three-component MN-FMR is chosen by the BIC, with a 6.67\% of misclassified units.
Therefore, our model is able to completely recognize the underlying group structure, differently from the MN-FMR.
Notice that, even if we consider the MN-FMR directly with $G=2$, this will produce a similar classification with an ARI of 0.89.

The estimated regression coefficients by the MN-CWM are displayed in \tablename~\ref{tab:Anvur_Coeff}, for the two groups, respectively.
\begin{table}[ht]
\caption{Education data: estimated regression coefficients for the MN-CWM.}
\centering
\subfloat[Subtable 1 list of tables text][Bachelor's degrees]{
\begin{tabular}{c|ccc}
\toprule
				  & Intercept & Credits & Students \\ 
				\hline
         Complete & -0.087 & 0.973   & 0.102 \\
				 Drop     &  0.727 & -0.557  & -0.178 \\
				\bottomrule
\end{tabular}}
\qquad
\subfloat[Subtable 2 list of tables text][Master's degrees]{
\begin{tabular}{c|ccc}
\toprule
         & Intercept & Credits & Students  \\  
				\hline
         Complete & 0.309 & 0.101 & 0.587  \\
				 Drop     & 0.049 & 0.044 & -0.011 \\
				 \bottomrule
\end{tabular}}
\label{tab:Anvur_Coeff}
\end{table}
As it is plausible to expect, both covariates have a positive effect on the successful completion of the study programs, even if their magnitude is different in the two groups.
The credits obtained during the first year of study might be more important for bachelor's students, considering the difficulties that arise in the transition from high schools to universities
\citep{krause2006being}.
At the same time, obtaining at least 40 course credits per year should be easier for masters' students, resulting in a greater importance for the completion of the studies.
Conversely, both covariates have a negative impact on the drop rates, with the exception of the Credits variable that surprisingly turns to have a positive sign for the master's courses.

\subsection{Insurance data}
\label{subsec:Insurance}

For this second real data application, the ``Insurance'' dataset included in the \textbf{splm} package \citep{splm} is used.
This dataset was introduced by \citet{millo2011} to study the consumption of non-life insurance across the $N=103$ Italian provinces in the years 1998--2002 ($r=5$).

In this application we select, as responses, the following $p=2$ variables that are related to the consumption and the presence of insurance products in the market
\begin{enumerate}
\item real per-capita non-life premiums in 2000 euros (PPCD),
\item density of insurance agencies per 1000 inhabitants (AGEN),
\end{enumerate} 
and the following $q=3$ financial covariates
\begin{enumerate}
\item real per-capita GDP (RGDP),
\item real per-capita bank deposits (BANK),
\item real interest rate on lending to families and small enterprises (RIRS).
\end{enumerate}
The reasons why we focused on this subset of covariates are: (1) they are almost regularly used in the literature, and their relevant effects on the consumption or development of insurance products has been widely discussed \citep[see the references in][for further details]{millo2011}. Indeed, they are commonly used as proxies for income and general level of economic activity (RGDP), stock of wealth (BANK) and opportunity cost of allocate funds in insurance policies (RIRS); (2) avoid an excessive parametrization of the models.   
Notice that, for a better interpretation of the regression coefficients, the variables RGDP and BANK are divided by 1,000; thus, thousand of euros are considered.

\begin{figure}[!ht]
\centering
\subfloat[]{%
\resizebox*{7.5cm}{!}{\includegraphics{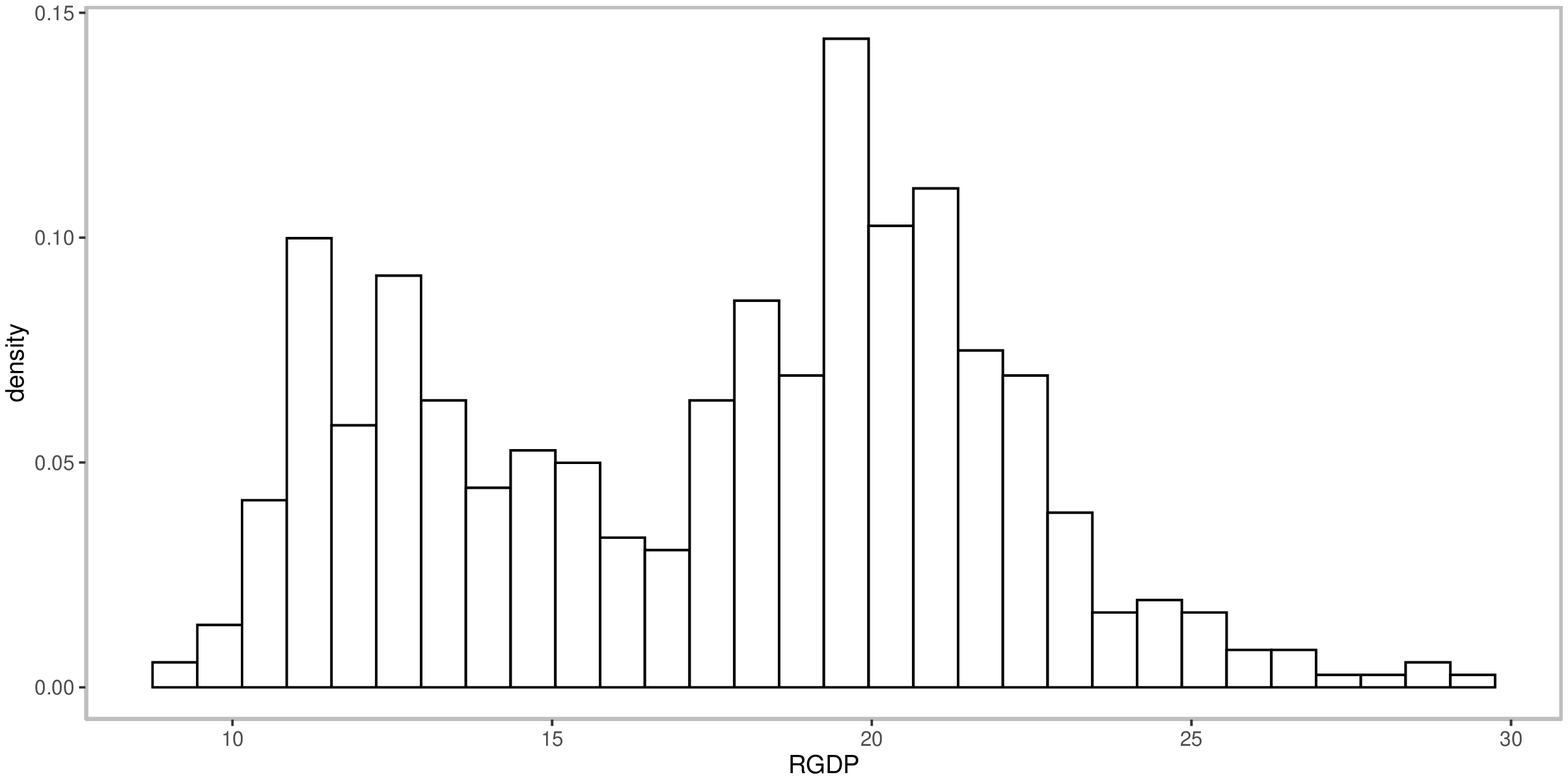}}}\hspace{5pt}
\subfloat[]{%
\resizebox*{7.5cm}{!}{\includegraphics{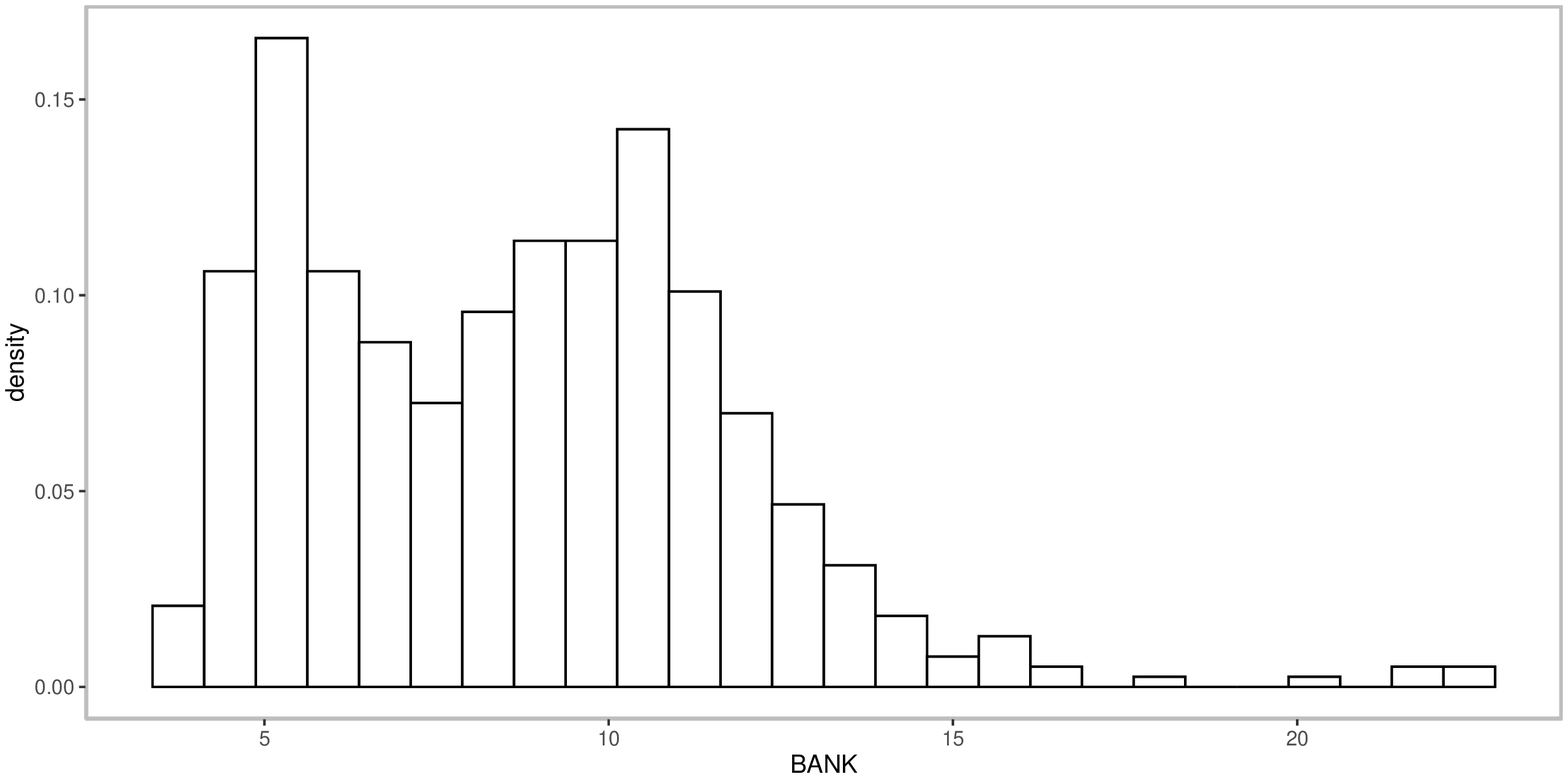}}}\hspace{5pt}
\subfloat[]{%
\resizebox*{7.5cm}{!}{\includegraphics{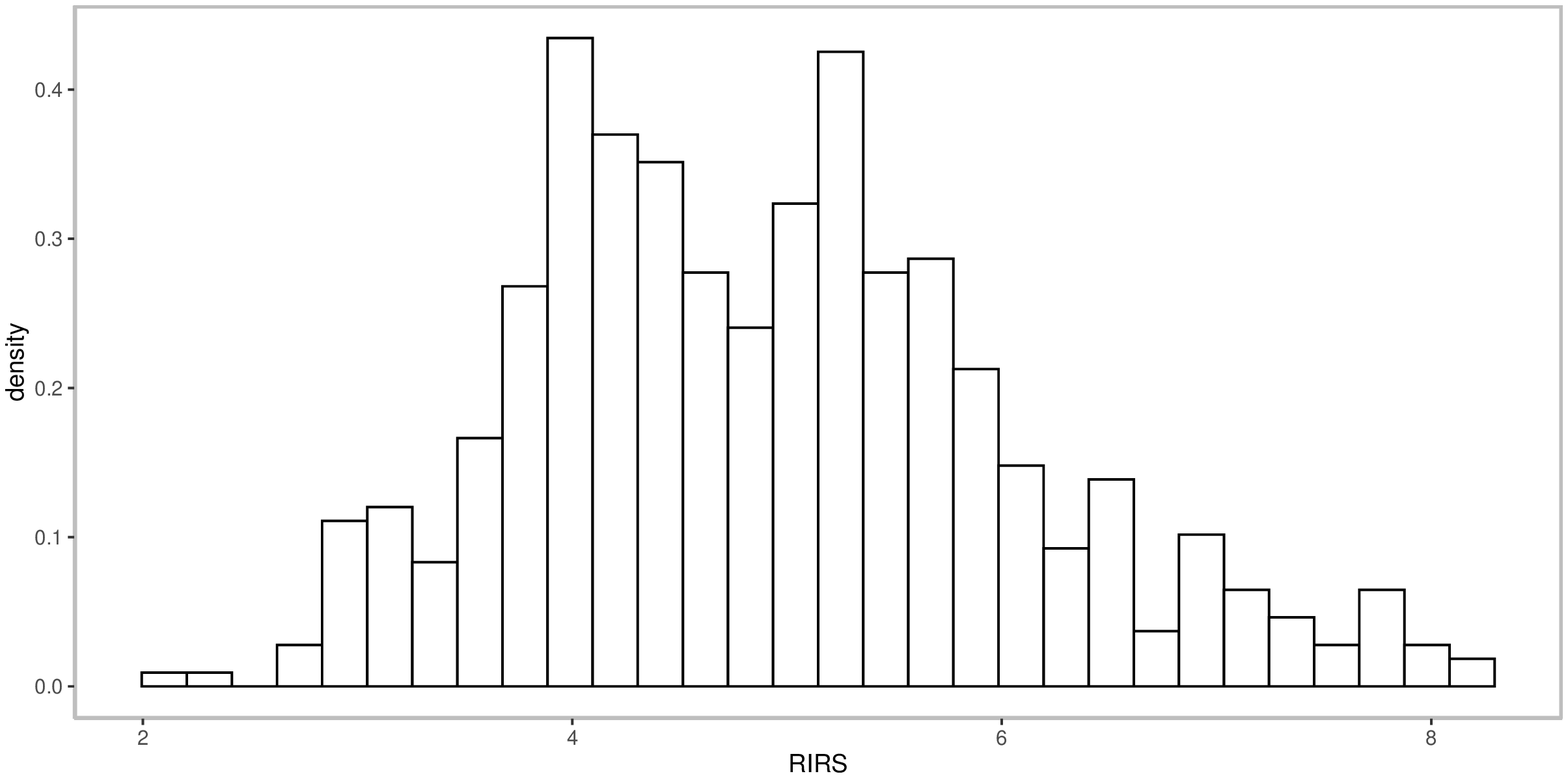}}}\hspace{5pt}
\caption{Sampling distributions of the covariates.}
\label{fig:hist}
\end{figure}

\begin{figure}[!ht]
\centering
\subfloat[]{%
\resizebox*{7.5cm}{!}{\includegraphics{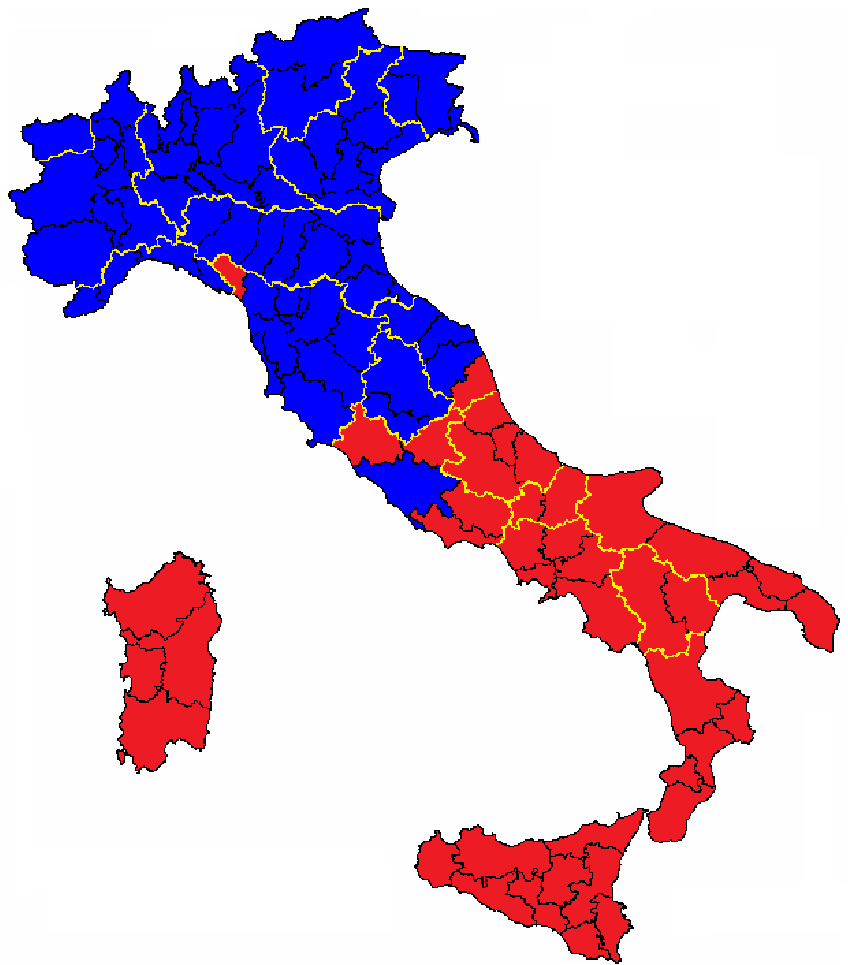}}}\hspace{5pt}
\subfloat[]{%
\resizebox*{7.5cm}{!}{\includegraphics{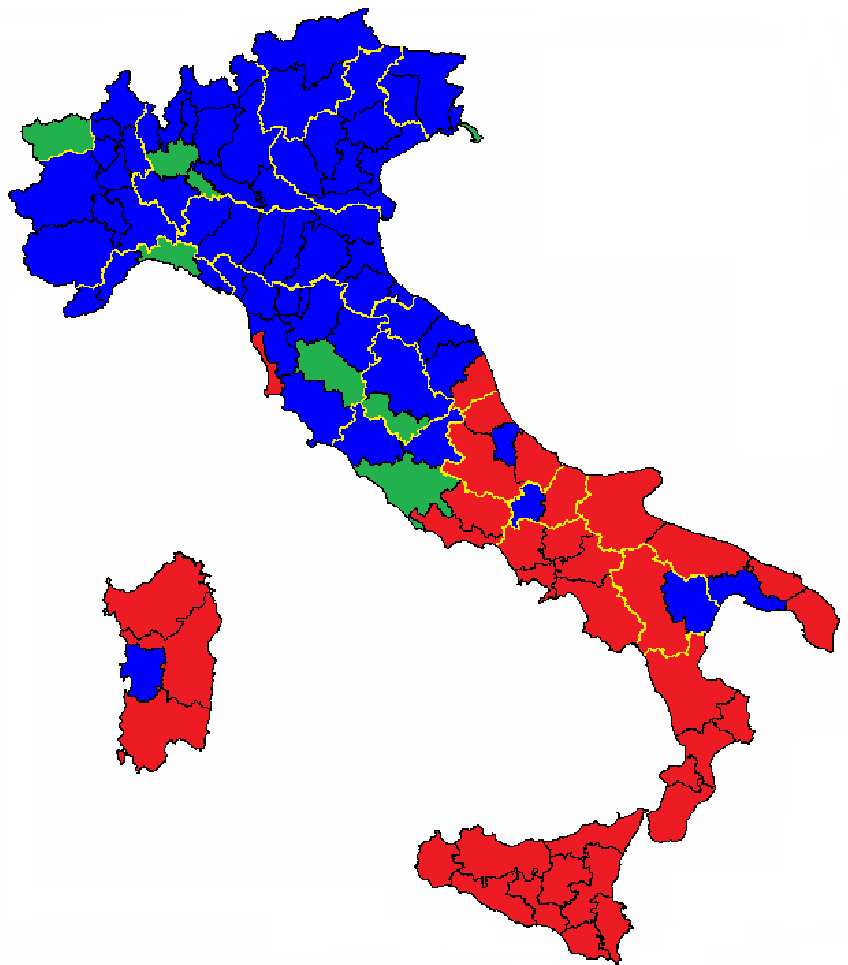}}}\hspace{5pt}
\caption{Partitions produced by the MN-CWM (a) and MN-FMR (b).}
\label{fig:ita}
\end{figure}

Differently from the previous application, we don't have a classification of the data.
However, since we are using $p\left(\bX\right)$ in~\eqref{eq:NOR_CWM}, the sampling distribution of each covariate could provide useful insights.
They are reported in \figurename~\ref{fig:hist}, for the full 5-year data, as done by \citet{melnykov2019studying}.
The bimodality in all these histograms seems to suggest the presence of two groups in the data, validating the findings of \citet{millo2011}.
They underline the existence of two macro areas, namely the Central-Northern Italy, characterized by an insurance penetration level relatively close to the European averages, and the South Italy, where a general economic underdevelopment has long been standing as a fundamental social and political problem.
The MN-CWM and the MN-FMR models are hence fitted to the data for $G\in\left\{1,2,3\right\}$, and the BIC selects a two-component MN-CWM and a three-component MN-FMR, respectively. 
To give a representation of these two partitions, they are illustrated in \figurename~\ref{fig:ita} by using the Italian political map.
Specifically, the Italian regions are bordered in yellow (islands excluded), while the internal provinces are delimited with the black lines and colored according to the estimated group membership both for the MN-CWM and the MN-FMR.
Since we do not have a classification of the data, we cannot compute neither ARI nor $\eta$.
Nevertheless, the partition produced by the MN-CWM seems in line with the findings of \citet{millo2011}, with a clear separation between the Central-Northern Italy and the Southern-Insular Italy.
Furthermore, with the exclusion of three cases, all the provinces belonging to the same region are clustered together.
The only exceptions concern the province of Rome (in the Lazio region), which due to its social-economic development is reasonably assigned to the Central-Northern Italy group, the province of Ascoli-Piceno (in the Marche region) and the province of Massa-Carrara (in the Toscana region).
On the contrary, the three groups detected by the MN-FRM are not supported by the literature and are difficult to interpret, even because they put together provinces spanning all over the country without a straightforward and reasonable justification.

As in Section~\ref{subsec:ANVUR}, the estimated regression coefficients by the MN-CWM are briefly presented in \tablename~\ref{tab:ins_Coeff}, for the two groups. 
\begin{table}[!ht]
\caption{Insurance data: estimated regression coefficients by the MN-CWM}
\centering
\subfloat[Subtable 1 list of tables text][Central-Northern Italy]{
\begin{tabular}{l|cccc}
\toprule
     & Intercept & RGDP & BANK & RIRS \\
    \hline
    PPCD & 85.7936 & 9.1932 & 1.8513 & -7.3079 \\
		AGEN & 0.5343  & -0.0085 & 0.0071 & 0.0073 \\
    \bottomrule\end{tabular}}
\qquad
\subfloat[Subtable 2 list of tables text][South Italy]{
\begin{tabular}{l|cccc}
\toprule
     & Intercept & RGDP & BANK & RIRS \\
				\hline
         PPCD &  -3.6968 & 6.0029 & 4.2062 & -1.2885 \\
				 AGEN & 0.0307 & 0.0041 & 0.0279  & 0.0039 \\
    \bottomrule\end{tabular}}
\label{tab:ins_Coeff}
\end{table}
Overall, the coefficients are quite different between the two groups, highlighting the divergences that characterize these two macro areas.
The RIRS has a negative impact on the PPCD, because the higher the interest rate on lending, the higher the opportunity cost of investing in insurance products.
Conversely, an increase in the RIRS has a positive effect on the AGEN because, from the insurance companies point of view, it rises the gains of investing the premiums on financial markets in the time between premium collection and claims settling.
Therefore, these increased revenues might lead to an expansion of the insurance companies over the territory.
Similarly, and in accordance with existing literature, income (RGDP) and wealth (BANK) have a positive influence on the PPCD \citep{millo2011}.
It would be reasonable to expect that they have a positive effect also on AGEN.
However, this is not the case at least for the Central-Northern Italy, that shows lower coefficients than the South Italy.
In the southern part of the country, where the insurance density and penetration are much lower, an increased income and wealth can give more space to growth and investment opportunities.

\section{Conclusion}
\label{sec:conclude}

The matrix-variate CWM has been introduced in this work.
In the MN-CWM framework, the sets of responses and covariates may be simultaneously observed at different time points or locations.
The matrix normal was used for both the cluster-specific distributions of covariates and responses given covariates, and an ECM algorithm for parameter estimation was presented.
A simulation study analyzed parameter recovery and the classification performance of the proposed model as well as the capability of the BIC to detect the underlying group structure of the data.
A comparison among the different initialization strategies has been also conducted.
Additionally, the MN-CWM has been compared to the MN-FMR, discussing under which scenarios the cluster-specific distributions of covariates is useful, and to the MMN-CWM on the vectorized data, where the overparameterization issue and its consequences in model selection have been analyzed.
Lastly, in the first real data application, the MN-CWM produced a perfect classification, differently from the MN-FMR model.
Similarly, in the second real data analysis, our model seemed to provide a more reliable partition of the Italian provinces than the MN-FMR model.

Further model developments can be readily proposed.
It is reasonable to assume that similar overparameterization concerns can affect also the MN-CWM when the dimensions of the matrices are quite high.
To further increase the parsimony of our model, constrained parameterizations of the covariance matrices can be employed, by following the approaches of \citet{sarkar2019parsimonious}, \citet{Sube:Punz:Ingr:McNi:Clus:2013}, \citet{Punz:Ingr:Pars:2015} and \citet{JSS:flexCWM}.
Furthermore, to accommodate for skewness or mild outliers in the data, skew or heavy tailed matrix-variate distributions could also be considered for the mixing components of the model \citep{melnykov2018model,gallaugher2018finite,tomarchio2020two}.

\begin{appendices}
\section{}
\label{sec:app1}
\begin{table}[H]
    \centering
		\caption{Parameters used to generate the simulated datasets under Scenario A$_{1}$.}
\label{tab:param1}       % Give a unique label
\resizebox{\columnwidth}{!}{%
     \begin{tabular}{lccccc}
				\hline\noalign{\smallskip}
Parameter  & Group 1 & Group 2 & Group 3 & Group 4\\
	      \hline\noalign{\smallskip}
	$\pi_g$	& 0.30 & 0.30 & 0.20 & 0.20 \\		
  $\bM_{g}$           & $\setlength\arraycolsep{2.8pt}\begin{pmatrix*}[r] 1.00 &  2.00 &  0.00 \\														
															                                       -4.00 & -3.00 & -3.00 \\
                                                                      1.00 &  2.00 &  1.00  \end{pmatrix*}$ 
								      & $\setlength\arraycolsep{2.8pt}\begin{pmatrix*}[r]  6.00 & 8.00 & 6.00 \\
                                                                       2.00 & 1.00 & 3.00  \\
                                                                       5.00 & 6.00 & 6.00 \end{pmatrix*}$	
											& $\setlength\arraycolsep{2.8pt}\begin{pmatrix*}[r]  -4.00 & -3.00 & -4.00 \\
                                                                       -9.00 & -9.00 & -7.00  \\
                                                                       -4.00 & -3.00 & -5.00  \end{pmatrix*}$																											
											& $\setlength\arraycolsep{2.8pt}\begin{pmatrix*}[r]  12.00 & 12.00 & 11.00 \\
                                                                        6.00 & 7.00  &  7.00  \\
                                                                       10.00 & 11.00 & 11.00  \end{pmatrix*}$ \\	
	$\bPhi_{\bX_{g}}$   & $\setlength\arraycolsep{2.8pt}\begin{pmatrix*}[r]	 1.00 & 0.50 & 0.25 \\    
                                                                       0.50 & 1.00 & 0.50 \\
                                                                       0.25 & 0.50 & 1.00 \end{pmatrix*}$ 
								      & $\setlength\arraycolsep{2.8pt}\begin{pmatrix*}[r]  2.00 & 0.40 & 0.08 \\
                                                                       0.40 & 0.20 & 0.40 \\
                                                                       0.08 & 0.40 & 2.00 \end{pmatrix*}$
											& $\setlength\arraycolsep{2.8pt}\begin{pmatrix*}[r]	 1.50 & 0.75 & 0.38 \\    
                                                                       0.75 & 1.50 & 0.75 \\
                                                                       0.38 & 0.75 & 1.50 \end{pmatrix*}$ 
								      & $\setlength\arraycolsep{2.8pt}\begin{pmatrix*}[r]  1.20 & 0.60 & 0.30 \\
                                                                       0.60 & 1.20 & 0.60 \\
                                                                       0.30 & 0.60 & 1.20 \end{pmatrix*}$\\
  $\bPsi_{\bX_{g}}$   & $\setlength\arraycolsep{2.8pt}\begin{pmatrix*}[r]  1.20 & 0.60 & 0.30 \\
                                                                       0.60 & 1.20 & 0.60 \\
																		                                   0.30 & 0.60 & 1.20 \end{pmatrix*}$ 
								      & $\setlength\arraycolsep{2.8pt}\begin{pmatrix*}[r]  1.40 & 0.70 & 0.35 \\
                                                                       0.70 & 1.40 & 0.70 \\
                                                                       0.35 & 0.70 & 1.40 \end{pmatrix*}$ 
											& $\setlength\arraycolsep{2.8pt}\begin{pmatrix*}[r]  0.80 & 0.40 & 0.20 \\
                                                                       0.40 & 0.80 & 0.40 \\
																		                                   0.20 & 0.40 & 0.80 \end{pmatrix*}$ 
								      & $\setlength\arraycolsep{2.8pt}\begin{pmatrix*}[r]  1.60 & 0.80 & 0.40 \\
                                                                       0.80 & 1.60 & 0.80 \\
                                                                       0.40 & 0.80 & 1.60 \end{pmatrix*}$\\
 $\bB^{*}_{g}$        & $\setlength\arraycolsep{2.8pt}\begin{pmatrix*}[r]  0.00 & 1.00 & 1.00 & 1.00 \\
                                                                      -2.00 & 1.00 & 1.50 & 1.00 \\
																																			 1.00 & 1.50 & 1.50 & 1.00 \end{pmatrix*}$   
								      & $\setlength\arraycolsep{2.8pt}\begin{pmatrix*}[r]  6.00 &-1.00 &-1.50 &-1.00 \\
			                                                                 4.00 &-1.00 &-1.50 &-1.00 \\
																																			 8.00 &-1.50 &-1.50 &-1.00 \end{pmatrix*}$
											& $\setlength\arraycolsep{2.8pt}\begin{pmatrix*}[r] -5.00 & 1.00 & 1.00 & 1.00 \\
			                                                                -3.00 & 1.50 & 1.00 & 1.00 \\
																																			-6.00 & 1.50 & 1.50 & 1.00 \end{pmatrix*}$		
											& $\setlength\arraycolsep{2.8pt}\begin{pmatrix*}[r]  1.00 &-1.00 &-1.00 &-1.00 \\
			                                                                -5.00 &-1.00 &-1.50 &-1.50 \\
																																			 0.00 &-1.50 &-1.00 &-1.50 \end{pmatrix*}$\\
$\bPhi_{\bY_{g}}$     & $\setlength\arraycolsep{2.8pt}\begin{pmatrix*}[r]  1.40 & 0.84 & 0.50 \\
                                                                       0.84 & 1.40 & 0.84 \\
                                                                       0.50 & 0.84 & 1.40 \end{pmatrix*}$ 
								      &	$\setlength\arraycolsep{2.8pt}\begin{pmatrix*}[r]  1.80 & 1.26 & 0.88 \\
                                                                       1.26 & 1.80 & 1.26 \\
                                                                       0.88 & 1.26 & 1.80 \end{pmatrix*}$
											& $\setlength\arraycolsep{2.8pt}\begin{pmatrix*}[r]  1.20 & 0.84 & 0.59 \\
                                                                       0.84 & 1.20 & 0.84 \\
                                                                       0.59 & 0.84 & 1.20 \end{pmatrix*}$ 
								      &	$\setlength\arraycolsep{2.8pt}\begin{pmatrix*}[r]  1.60 & 0.96 & 0.58 \\
                                                                       0.96 & 1.60 & 0.96 \\
                                                                       0.58 & 0.96 & 1.60 \end{pmatrix*}$\\
$\bPsi_{\bY_{g}}$     & $\setlength\arraycolsep{2.8pt}\begin{pmatrix*}[r]  2.00 & 0.60 & 0.18 \\
                                                                       0.60 & 0.20 & 0.60 \\
                                                                       0.18 & 0.60 & 2.00 \end{pmatrix*}$ 
								      & $\setlength\arraycolsep{2.8pt}\begin{pmatrix*}[r]  1.10 & 0.55 & 0.28 \\
                                                                       0.55 & 1.10 & 0.55 \\
                                                                       0.28 & 0.55 & 1.10 \end{pmatrix*}$
											& $\setlength\arraycolsep{2.8pt}\begin{pmatrix*}[r]  1.90 & 1.71 & 1.54 \\
                                                                       1.71 & 1.90 & 1.71 \\
                                                                       1.54 & 1.71 & 1.90 \end{pmatrix*}$ 
								      & $\setlength\arraycolsep{2.8pt}\begin{pmatrix*}[r]  1.40 & 1.26 & 1.13 \\
                                                                       1.26 & 1.40 & 1.26 \\
                                                                       1.13 & 1.26 & 1.40 \end{pmatrix*}$\\
\noalign{\smallskip}\hline
        \end{tabular}%
}
\end{table}
\clearpage
\section{}
\label{sec:app2}
\begin{table}[!h]
    \centering
		\caption{Parameters used to generate the simulated datasets under Scenario A$_{2}$.}
\label{tab:param2}       % Give a unique label
\resizebox{0.60\columnwidth}{!}{%
     \begin{tabular}{lcc}
				\hline\noalign{\smallskip}
Parameter  & Group 1 & Group 2 \\
	      \hline\noalign{\smallskip}
	$\pi_g$	& 0.50 & 0.50  \\		
  $\bM_{g}$           & $\setlength\arraycolsep{2.8pt}\begin{pmatrix*}[r] 1.00 & 2.00 & 2.00 & 0.00 \\														
															                                           -1.00 & 1.00 & 1.00 & 2.00 \\
                                                                          0.00 & 2.00 & 2.00 & 1.00 \end{pmatrix*}$ 
								      & $\setlength\arraycolsep{2.8pt}\begin{pmatrix*}[r] 1.00 & 2.00 & 2.00 & 0.00 \\														
															                                           -1.00 & 1.00 & 1.00 & 2.00 \\
                                                                          0.00 & 2.00 & 2.00 & 1.00 \end{pmatrix*}$ \\
	$\bPhi_{\bX_{g}}$   & $\setlength\arraycolsep{2.8pt}\begin{pmatrix*}[r]	 1.00 & 0.50 & 0.25 \\    
                                                                           0.50 & 1.00 & 0.50 \\
                                                                           0.25 & 0.50 & 1.00 \end{pmatrix*}$ 
								      & $\setlength\arraycolsep{2.8pt}\begin{pmatrix*}[r]  2.00 & 0.40 & 0.08 \\
                                                                           0.40 & 0.20 & 0.40 \\
                                                                           0.08 & 0.40 & 2.00 \end{pmatrix*}$ \\
  $\bPsi_{\bX_{g}}$   & $\setlength\arraycolsep{2.8pt}\begin{pmatrix*}[r]  1.70 & 0.85 & 0.42 & 0.21\\
                                                                           0.85 & 1.70 & 0.85 & 0.42\\
																		                                       0.42 & 0.85 & 1.70 & 0.85\\
																																					 0.21 & 0.42 & 0.85 & 1.70 \end{pmatrix*}$ 
								      & $\setlength\arraycolsep{2.8pt}\begin{pmatrix*}[r]  1.00 & 0.50 & 0.25 & 0.12 \\
                                                                           0.50 & 1.00 & 0.50 & 0.25 \\
                                                                           0.25 & 0.50 & 1.00 & 0.50 \\
																																					 0.12 & 0.25 & 0.50 & 1.00 \end{pmatrix*}$ \\
 $\bB^{*}_{g}$        & $\setlength\arraycolsep{2.8pt}\begin{pmatrix*}[r]  2.00 & 1.00 & 1.00 &-1.00 \\
                                                                           3.00 & 1.00 &-1.00 & 1.00 \end{pmatrix*}$   
								      & $\setlength\arraycolsep{2.8pt}\begin{pmatrix*}[r] -7.00 & 1.00 & 1.00 &-1.00 \\
			                                                                    -8.00 & 1.00 &-1.00 & 1.00 \end{pmatrix*}$\\
$\bPhi_{\bY_{g}}$     & $\setlength\arraycolsep{2.8pt}\begin{pmatrix*}[r]  1.00 & 0.50  \\
                                                                           0.50 & 1.00  \end{pmatrix*}$ 
								      &	$\setlength\arraycolsep{2.8pt}\begin{pmatrix*}[r]  2.00 & 1.20  \\
                                                                           1.20 & 2.00  \end{pmatrix*}$ \\
$\bPsi_{\bY_{g}}$     & $\setlength\arraycolsep{2.8pt}\begin{pmatrix*}[r]  2.00 & 1.00 & 0.50 & 0.25 \\
                                                                           1.00 & 2.00 & 1.00 & 0.50 \\
                                                                           0.50 & 1.00 & 2.00 & 1.00 \\
																																					 0.25 & 0.50 & 1.00 & 2.00 \end{pmatrix*}$ 
								      & $\setlength\arraycolsep{2.8pt}\begin{pmatrix*}[r]  1.70 & 0.75 & 0.38 & 0.19 \\
                                                                           0.75 & 1.50 & 0.75 & 0.38 \\
                                                                           0.38 & 0.75 & 1.50 & 0.75 \\
																																					 0.19 & 0.39 & 0.75 & 1.50 \end{pmatrix*}$ \\
\noalign{\smallskip}\hline
        \end{tabular}%
}
\end{table}
\end{appendices}

\bibliographystyle{chicago}

\bibliography{Bibliography-MM-MC}

\begin{thebibliography}{}

\bibitem[\protect\citeauthoryear{Anderlucci, Montanari, and Viroli}{Anderlucci
  et~al.}{2014}]{anderlucci2014matrix}
Anderlucci, L., A.~Montanari, and C.~Viroli (2014).
\newblock A matrix-variate regression model with canonical states: An
  application to elderly danish twins.
\newblock {\em Statistica\/}~{\em 74\/}(4), 367--381.

\bibitem[\protect\citeauthoryear{Dang and McNicholas}{Dang and
  McNicholas}{2015}]{dang2015families}
Dang, U.~J. and P.~D. McNicholas (2015).
\newblock Families of parsimonious finite mixtures of regression models.
\newblock In {\em Advances in Statistical Models for Data Analysis}, pp.\
  73--84. Springer.

\bibitem[\protect\citeauthoryear{Dang, Punzo, McNicholas, Ingrassia, and
  Browne}{Dang et~al.}{2017}]{dang2017multivariate}
Dang, U.~J., A.~Punzo, P.~D. McNicholas, S.~Ingrassia, and R.~P. Browne (2017).
\newblock Multivariate response and parsimony for {G}aussian cluster-weighted
  models.
\newblock {\em Journal of Classification\/}~{\em 34\/}(1), 4--34.

\bibitem[\protect\citeauthoryear{Dayton and Macready}{Dayton and
  Macready}{1988}]{dayton1988concomitant}
Dayton, C.~M. and G.~B. Macready (1988).
\newblock Concomitant-variable latent-class models.
\newblock {\em Journal of the American Statistical Association\/}~{\em
  83\/}(401), 173--178.

\bibitem[\protect\citeauthoryear{Dempster, Laird, and Rubin}{Dempster
  et~al.}{1977}]{dempster1977maximum}
Dempster, A.~P., N.~M. Laird, and D.~B. Rubin (1977).
\newblock Maximum likelihood from incomplete data via the em algorithm.
\newblock {\em Journal of the Royal Statistical Society: Series B
  (Methodological)\/}~{\em 39\/}(1), 1--22.

\bibitem[\protect\citeauthoryear{DeSarbo and Cron}{DeSarbo and
  Cron}{1988}]{desarbo1988}
DeSarbo, W.~S. and W.~L. Cron (1988).
\newblock A maximum likelihood methodology for clusterwise linear regression.
\newblock {\em Journal of Classification\/}~{\em 5\/}(2), 249--282.

\bibitem[\protect\citeauthoryear{Dutilleul}{Dutilleul}{1999}]{dutilleul1999mle}
Dutilleul, P. (1999).
\newblock The {MLE} algorithm for the matrix normal distribution.
\newblock {\em Journal of Statistical Computation and Simulation\/}~{\em
  64\/}(2), 105--123.

\bibitem[\protect\citeauthoryear{Fr{\"u}hwirth-Schnatter}{Fr{\"u}hwirth-Schnatter}{2006}]{fruhwirth2006finite}
Fr{\"u}hwirth-Schnatter, S. (2006).
\newblock {\em Finite mixture and Markov switching models}.
\newblock Springer Science \& Business Media.

\bibitem[\protect\citeauthoryear{Gallaugher and McNicholas}{Gallaugher and
  McNicholas}{2018}]{gallaugher2018finite}
Gallaugher, M.~P. and P.~D. McNicholas (2018).
\newblock Finite mixtures of skewed matrix variate distributions.
\newblock {\em Pattern Recognition\/}~{\em 80}, 83--93.

\bibitem[\protect\citeauthoryear{Gallaugher and McNicholas}{Gallaugher and
  McNicholas}{2017}]{gallaugher17}
Gallaugher, M. P.~B. and P.~D. McNicholas (2017).
\newblock A matrix variate skew-t distribution.
\newblock {\em Stat\/}~{\em 6\/}(1), 160--170.

\bibitem[\protect\citeauthoryear{Gershenfeld}{Gershenfeld}{1997}]{gershenfeld1997}
Gershenfeld, N. (1997).
\newblock Nonlinear inference and cluster-weighted modeling.
\newblock {\em Annals of the New York Academy of Sciences\/}~{\em 808\/}(1),
  18--24.

\bibitem[\protect\citeauthoryear{Gershenfeld, Schoner, and Metois}{Gershenfeld
  et~al.}{1999}]{gershenfeld1999}
Gershenfeld, N., B.~Schoner, and E.~Metois (1999).
\newblock Cluster-weighted modelling for time-series analysis.
\newblock {\em Nature\/}~{\em 397\/}(6717), 329.

\bibitem[\protect\citeauthoryear{Hubert and Arabie}{Hubert and
  Arabie}{1985}]{hubert1985comparing}
Hubert, L. and P.~Arabie (1985).
\newblock Comparing partitions.
\newblock {\em Journal of Classification\/}~{\em 2\/}(1), 193--218.

\bibitem[\protect\citeauthoryear{Ingrassia, Minotti, and Punzo}{Ingrassia
  et~al.}{2014}]{ingrassia2014model}
Ingrassia, S., S.~C. Minotti, and A.~Punzo (2014).
\newblock Model-based clustering via linear cluster-weighted models.
\newblock {\em Computational Statistics \& Data Analysis\/}~{\em 71}, 159--182.

\bibitem[\protect\citeauthoryear{Ingrassia, Minotti, and Vittadini}{Ingrassia
  et~al.}{2012}]{ingrassia2012local}
Ingrassia, S., S.~C. Minotti, and G.~Vittadini (2012).
\newblock Local statistical modeling via a cluster-weighted approach with
  elliptical distributions.
\newblock {\em Journal of Classification\/}~{\em 29\/}(3), 363--401.

\bibitem[\protect\citeauthoryear{Ingrassia and Punzo}{Ingrassia and
  Punzo}{2016}]{ingrassia2016decision}
Ingrassia, S. and A.~Punzo (2016).
\newblock Decision boundaries for mixtures of regressions.
\newblock {\em Journal of the Korean Statistical Society\/}~{\em 45\/}(2),
  295--306.

\bibitem[\protect\citeauthoryear{Ingrassia and Punzo}{Ingrassia and
  Punzo}{2020}]{ingrassia2019cluster}
Ingrassia, S. and A.~Punzo (2020).
\newblock Cluster validation for mixtures of regressions via the total sum of
  squares decomposition.
\newblock {\em Journal of Classification\/}~{\em 37\/}(2), 526--547.

\bibitem[\protect\citeauthoryear{Ingrassia, Punzo, Vittadini, and
  Minotti}{Ingrassia et~al.}{2015}]{ingrassia2015generalized}
Ingrassia, S., A.~Punzo, G.~Vittadini, and S.~C. Minotti (2015).
\newblock The generalized linear mixed cluster-weighted model.
\newblock {\em Journal of Classification\/}~{\em 32\/}(1), 85--113.

\bibitem[\protect\citeauthoryear{Krause}{Krause}{2006}]{krause2006being}
Krause, K. (2006).
\newblock On being strategic about the first year.
\newblock In {\em Keynote presentation, Queensland University of Technology
  First Year Forum}, Volume~5.

\bibitem[\protect\citeauthoryear{Leisch}{Leisch}{2004}]{leisch2004flexmix}
Leisch, F. (2004).
\newblock \textbf{FlexMix}: A general framework for finite mixture models and
  latent class regression in \textsf{R}.
\newblock {\em Journal of Statistical Software\/}~{\em 11\/}(8), 1--18.

\bibitem[\protect\citeauthoryear{Maitra and Melnykov}{Maitra and
  Melnykov}{2010}]{maitra2010simulating}
Maitra, R. and V.~Melnykov (2010).
\newblock Simulating data to study performance of finite mixture modeling and
  clustering algorithms.
\newblock {\em Journal of Computational and Graphical Statistics\/}~{\em
  19\/}(2), 354--376.

\bibitem[\protect\citeauthoryear{Mazza, Battisti, Ingrassia, and Punzo}{Mazza
  et~al.}{2019}]{Mazz:Batt:Ingr:Punz:Mode:2019}
Mazza, A., M.~Battisti, S.~Ingrassia, and A.~Punzo (2019).
\newblock Modeling return to education in heterogeneous populations. an
  application to italy.
\newblock In F.~Greselin, L.~Deldossi, L.~Bagnato, and M.~Vichi (Eds.), {\em
  Statistical Learning of Complex Data}, Volume~88 of {\em Studies in
  Classification, Data Analysis, and Knowledge Organization}, Cham,
  Switzerland, pp.\  121--131. Springer.

\bibitem[\protect\citeauthoryear{Mazza, Punzo, and Ingrassia}{Mazza
  et~al.}{2018}]{JSS:flexCWM}
Mazza, A., A.~Punzo, and S.~Ingrassia (2018).
\newblock \textbf{flexCWM}: A flexible framework for cluster-weighted models.
\newblock {\em Journal of Statistical Software\/}~{\em 86\/}(2), 1--30.

\bibitem[\protect\citeauthoryear{McLachlan and Peel}{McLachlan and
  Peel}{2000}]{mclachlan2000finite}
McLachlan, G. and D.~Peel (2000).
\newblock {\em Finite Mixture Models}.
\newblock John Wiley \& Sons.

\bibitem[\protect\citeauthoryear{McNicholas}{McNicholas}{2016a}]{mcnicholas2016mixture}
McNicholas, P.~D. (2016a).
\newblock {\em Mixture model-based classification}.
\newblock Chapman and Hall/CRC.

\bibitem[\protect\citeauthoryear{McNicholas}{McNicholas}{2016b}]{mcnicholas16b}
McNicholas, P.~D. (2016b).
\newblock Model-based clustering.
\newblock {\em Journal of Classification\/}~{\em 33\/}(3), 331--373.

\bibitem[\protect\citeauthoryear{Melnykov and Zhu}{Melnykov and
  Zhu}{2018}]{melnykov2018model}
Melnykov, V. and X.~Zhu (2018).
\newblock On model-based clustering of skewed matrix data.
\newblock {\em Journal of Multivariate Analysis\/}~{\em 167}, 181--194.

\bibitem[\protect\citeauthoryear{Melnykov and Zhu}{Melnykov and
  Zhu}{2019}]{melnykov2019studying}
Melnykov, V. and X.~Zhu (2019).
\newblock Studying crime trends in the usa over the years 2000--2012.
\newblock {\em Advances in Data Analysis and Classification\/}~{\em 13\/}(1),
  325--341.

\bibitem[\protect\citeauthoryear{Meng and Van~Dyk}{Meng and
  Van~Dyk}{1997}]{meng1997algorithm}
Meng, X.-L. and D.~Van~Dyk (1997).
\newblock The em algorithm—an old folk-song sung to a fast new tune.
\newblock {\em Journal of the Royal Statistical Society: Series B (Statistical
  Methodology)\/}~{\em 59\/}(3), 511--567.

\bibitem[\protect\citeauthoryear{Michael and Melnykov}{Michael and
  Melnykov}{2016}]{michael2016effective}
Michael, S. and V.~Melnykov (2016).
\newblock An effective strategy for initializing the em algorithm in finite
  mixture models.
\newblock {\em Advances in Data Analysis and Classification\/}~{\em 10\/}(4),
  563--583.

\bibitem[\protect\citeauthoryear{Millo and Carmeci}{Millo and
  Carmeci}{2011}]{millo2011}
Millo, G. and G.~Carmeci (2011).
\newblock Non-life insurance consumption in {I}taly: a sub-regional panel data
  analysis.
\newblock {\em Journal of Geographical Systems\/}~{\em 13\/}(3), 273--298.

\bibitem[\protect\citeauthoryear{Millo and Piras}{Millo and Piras}{2012}]{splm}
Millo, G. and G.~Piras (2012).
\newblock \textbf{splm}: Spatial panel data models in \textsf{R}.
\newblock {\em Journal of Statistical Software\/}~{\em 47\/}(1), 1--38.

\bibitem[\protect\citeauthoryear{Punzo}{Punzo}{2014}]{punzo2014flexible}
Punzo, A. (2014).
\newblock Flexible mixture modelling with the polynomial {G}aussian
  cluster-weighted model.
\newblock {\em Statistical Modelling\/}~{\em 14\/}(3), 257--291.

\bibitem[\protect\citeauthoryear{Punzo and Ingrassia}{Punzo and
  Ingrassia}{2015}]{Punz:Ingr:Pars:2015}
Punzo, A. and S.~Ingrassia (2015).
\newblock Parsimonious generalized linear {G}aussian cluster-weighted models.
\newblock In I.~Morlini, T.~Minerva, and M.~Vichi (Eds.), {\em Advances in
  Statistical Models for Data Analysis}, Studies in Classification, Data
  Analysis and Knowledge Organization, Switzerland, pp.\  201--209. Springer
  International Publishing.

\bibitem[\protect\citeauthoryear{Punzo and Ingrassia}{Punzo and
  Ingrassia}{2016}]{punzo2016clustering}
Punzo, A. and S.~Ingrassia (2016).
\newblock Clustering bivariate mixed-type data via the cluster-weighted model.
\newblock {\em Computational Statistics\/}~{\em 31\/}(3), 989--1013.

\bibitem[\protect\citeauthoryear{Punzo and McNicholas}{Punzo and
  McNicholas}{2017}]{punzo2017robust}
Punzo, A. and P.~D. McNicholas (2017).
\newblock Robust clustering in regression analysis via the contaminated
  {G}aussian cluster-weighted model.
\newblock {\em Journal of Classification\/}~{\em 34\/}(2), 249--293.

\bibitem[\protect\citeauthoryear{Qiu and Joe.}{Qiu and Joe.}{2015}]{clusterG}
Qiu, W. and H.~Joe. (2015).
\newblock {\em clusterGeneration: Random Cluster Generation (with Specified
  Degree of Separation)}.
\newblock R package version 1.3.4.

\bibitem[\protect\citeauthoryear{{R Core Team}}{{R Core
  Team}}{2018}]{R_software}
{R Core Team} (2018).
\newblock {\em R: A Language and Environment for Statistical Computing}.
\newblock Vienna, Austria: R Foundation for Statistical Computing.

\bibitem[\protect\citeauthoryear{Sarkar, Zhu, Melnykov, and Ingrassia}{Sarkar
  et~al.}{2019}]{sarkar2019parsimonious}
Sarkar, S., X.~Zhu, V.~Melnykov, and S.~Ingrassia (2019).
\newblock On parsimonious models for modeling matrix data.
\newblock {\em Computational Statistics \& Data Analysis\/}, 106822.

\bibitem[\protect\citeauthoryear{Schwarz et~al.}{Schwarz
  et~al.}{1978}]{schwarz1978estimating}
Schwarz, G. et~al. (1978).
\newblock Estimating the dimension of a model.
\newblock {\em The Annals of Statistics\/}~{\em 6\/}(2), 461--464.

\bibitem[\protect\citeauthoryear{Subedi, Punzo, Ingrassia, and
  McNicholas}{Subedi et~al.}{2013}]{Sube:Punz:Ingr:McNi:Clus:2013}
Subedi, S., A.~Punzo, S.~Ingrassia, and P.~D. McNicholas (2013).
\newblock Clustering and classification via cluster-weighted factor analyzers.
\newblock {\em Advances in Data Analysis and Classification\/}~{\em 7\/}(1),
  5--40.

\bibitem[\protect\citeauthoryear{Tomarchio, Punzo, and Bagnato}{Tomarchio
  et~al.}{2020}]{tomarchio2020two}
Tomarchio, S.~D., A.~Punzo, and L.~Bagnato (2020).
\newblock Two new matrix-variate distributions with application in model-based
  clustering.
\newblock {\em Computational Statistics \& Data Analysis\/}~{\em 152}, 107050.

\bibitem[\protect\citeauthoryear{Viroli}{Viroli}{2012}]{viroli2012matrix}
Viroli, C. (2012).
\newblock On matrix-variate regression analysis.
\newblock {\em Journal of Multivariate Analysis\/}~{\em 111}, 296--309.

\end{thebibliography}
\end{document}